\documentclass[12pt]{article}

\usepackage{amsmath,amssymb}
\usepackage{graphicx}
\usepackage{caption}
\usepackage{color}
\usepackage{dcolumn}
\usepackage{bm}
\usepackage[numbers,super,comma,sort&compress]{natbib}
\usepackage{authblk}

\definecolor{background-color}{gray}{0.98}

\title{Generalized Sturmian Functions in prolate spheroidal coordinates}

\author[1]{D. M. Mitnik}
\author[1]{F.A. L\'opez}
\author[2]{L. U. Ancarani}
\affil[1]{Instituto de Astronom\'{\i}a y F\'{\i}sica del
Espacio (IAFE),  CONICET-UBA,
C.C. 67, Suc. 28, (C1428EGA) Buenos Aires, Argentina.}
\affil[2]{Universit\'e de Lorraine, CNRS, LPCT, 57000 Metz, France.}

\begin{document}

\maketitle

%%%%%%%%%%%%%%%%%%%%%%%%%%%%%%%%%%%%%%%%%%%%%%%%%%%%%%%%%%%%%%%%%%%%%%
\begin{abstract}
With the aim of describing bound and continuum states for diatomic
molecules, we develop and implement a spectral method that makes
use of Generalized Sturmian Functions (GSF) in prolate spheroidal
coordinates. In order to master all computational issues,
we apply here the method to one--electron molecular ions and
compare it with benchmark data for both ground and excited states. 
We actually propose two different computational schemes to solve the
two coupled differential equations.

The first one is an iterative 1$d$ procedure in which one solves
alternately the angular and the radial equations, the latter
yielding the state energy. The second, named direct $2d$ method,
consists in representing the Hamiltonian matrix in a
two--dimensional GSF basis set, and its further diagonalization.
Both spectral schemes are timewise computationally efficient since
the basis elements are such that no derivatives have to be
calculated numerically. Moreover, very accurate results are
obtained with minimal basis sets. This is related on one side to
the use of the natural coordinate system and, on the other, to the
intrinsic good property of all GSF basis elements that are
constructed as to obey appropriate physical boundary conditions.
Compared to the iterative 1$d$ approach, the direct 2$d$ method is superior 
in the sense that several states are obtained simultaneously. 
However, if one is interested in a specific state, a better accuracy is 
achieved with the 1$d$ method using GSF generated specifically for that state.
The present implementation for bound states paves the way
for the study of continuum states involved in ionization of one or
two--electron diatomic targets.
 \end{abstract}

%\begin{keyword}
%Molecular Structure
%Generalized Sturmian Functions
%Spheroidal Prolates
%\end{keyword}

\maketitle

\pagebreak
%%%%%%%%%%%%%%%%%%%%%%%%%%%%%%%%%%%%%%%%%%%%%%%%%%%%%%%%%%%%%%%%%%%%%%
\section{INTRODUCTION}

The molecular ion H$_2^+$, as well as the isotopic forms such as
HD$^+$ or D$^+_2$, and other one--electron diatomics such as
HHe$^{+2}$ or HLi$^{+3}$, are the simplest molecular quantum
three-body problem with Coulomb interactions. H$_2^+$, in
particular, has been largely studied since the early days of
quantum mechanics \cite{Burrau,Hylleraas,Jaffe2}, and is presented
in standard molecular physics books as it allows one to understand
why molecules form. On top of being important in astrophysics (it
is involved in many reaction chains leading to the production of
polyatomic molecules), the molecular ion H$_2^+$ also serves as
benchmark to test any new molecular approach and numerical method.

In the fixed--nuclei approximation, it is well known that prolate
spheroidal coordinates make the Schr\"odinger equation separable
\cite{BransdenJoachain}. Aside from the simple azimuthal angle
dependence due to axial symmetry, the wavefunction depends on two
variables, one angular and one radial (actually quasi--angular and
quasi--radial). The H$_2^+$ bound structure can be found by
solving a system of two coupled ordinary differential equations,
one for each of these two variables. An analytical solution exists
formally \cite{Burrau,Hylleraas,Jaffe2} but involves two not so
tractable expansions and therefrom complicated energy equations
(see, \emph{e.g.}, \cite{Carrington} and references therein). In
practice, therefore, the energies are found numerically. This is
why a wide variety of methods, including iterative methods, have
been proposed and applied to solve the coupled equations. For
continuum states, necessary for example to describe ionization
processes from diatomic molecules, the energy is known and fixed.
However, these non--L$^2$ states are much more difficult to build
as they oscillate up to infinity. Some recent investigations
dedicated to their description in prolate spheroidal coordinates
include Ref. \cite{paperCN,Kereselidze2019}. Approximate single or
double continuum wavefunctions borrowed from the atomic literature
have been extended to the two--center case and employed to study
ionization processes
\cite{Serov2002,Chuka2004,Serov2005,Chuka2008}. Other approaches
consist in extending well established atomic numerical techniques
to the diatomic molecular case, using (see, \emph{e.g.},
\cite{Tao2009,SerovJoulakian2009}) or not using
 (see, \emph{e.g.}, \cite{Foster2007}) prolate spheroidal coordinates.

In the last decade, a spectral method named GSF has been developed
and implemented to study the structure of and scattering processes
on atomic systems \cite{Mitnik:11,Gasaneo:13}. The method uses
complete and orthogonal basis sets of Generalized Sturmian
Functions (GSF) with appropriate boundary conditions. Negative
energy GSFs allow one to study bound states. The helium atom, the
simplest atomic quantum three--body problem with Coulomb
interactions, served as a benchmark to put the method on solid
grounds, by studying in details convergence issues, the integrals
involved and the adequate choice of optimal parameters and
numerical packages (see \cite{Optimal} and references therein).
While the aim of the GSF method was not to compete with well
established structure codes, it proved to be very accurate at a
reduced computational cost because of intrinsic GSF properties  in
particular the adequate, and unique, asymptotic decay of all basis
elements.

After bound states, the GSF approach was rapidly implemented for
continuum states for which the good properties of positive energy
GSFs demonstrated the power of the method. Indeed, for continuum
states, the correct asymptotic behavior is crucial in any
scattering calculation as shown in applications to one and
two--electron atomic systems (see, \emph{e.g.},
\cite{HeDPI,Ambrosio2016, Ambrosio2017}). The method was first
presented in spherical coordinates, then extended to
hyperspherical coordinates but limited to atomic systems. An
extension to molecules with a heavy central nucleus has been
proposed in a one--center GSF approach \cite{GranadosPhD} and
applied to ionization processes
\cite{Granados2016,Granados2017,Ali2019}. Nothing, however, has
been proposed to deal with diatomic molecules.

The purpose of this manuscript is to develop and implement a GSF
method in prolate spheroidal coordinates, thus combing the two
advantages of (i) using the natural coordinates for diatomic
systems and (ii) the power of a spectral method together with the
intrinsically good GSF properties. The long term aim is to be able
to describe accurately single or double ionization of diatomic
molecules treated as a two--electron system. The development will
follow a path similar to the one adopted for the atomic case. We
will first consider bound one--electron molecules before moving to
the continuum part of the spectrum. By studying benchmark
one--electron molecular ions, such as the H$_2^+$, we wish to
validate the new computational procedure and code, check
thoroughly all convergence and precision issues, and test
the robustness with respect to the variation of the internuclear
distance.

We actually present here two distinct computational methods that 
serve different purposes.
In the first one, we adopt an iterative approach, solving alternately
the separated Schr\"odinger equations for the angular part and for
the radial part. This {\it iterative $1d$} procedure, which is
repeated until convergence, presents the novelty of using GSF with
appropriate boundary conditions. Because of such property the
approach results to be computationally efficient as only small
basis are needed to obtain very good energy levels. It is also
efficient in computing time because the GSF basis elements already
solve the Hamiltonian differential operator so that no derivative
calculation is needed at each iteration.
The present study allows us to establish the capability of the 
approach and master the related parameters when using appropriate GSF 
in prolate spheroidal coordinates.
The iterative $1d$ procedure puts the focus on the energy and wave function 
of a single molecular state.
The second method, called here the {\it direct $2d$} method, 
has a different scope since it provides a set of states at 
the same time.
It consists in representing the Hamiltonian matrix in a 
two--dimensional GSF basis set, and its
further diagonalization. On top of the same advantages as the
first method, the $2d$ spectral approach demonstrates its full
power by providing accurately many states simultaneously, and this
with very small basis.

The remainder of this paper is as follows. In Sec.~\ref{sec:theory} we
provide the theoretical framework of the proposed GSF method in
prolate spheroidal coordinates. Then in Sec.~\ref{sec:results} we
apply it to the ground and first three excited states of symmetric
(H$_2^+$) and asymmetric (HHe$^{+2}$ and HLi$^{+3}$) molecular
ions. The successful comparison with benchmark data from the
literature allows us to validate the method for bound states. As
indicated in the Conclusion (Sec.~\ref{sec:conclusion}),
 the next step will be to
study continuum states for which positive energy GSF, with
appropriate boundary conditions, will be used.

Atomic units ($\hbar=m_e=e=1$) are assumed throughout.

%%%%%%%%%%%%%%%%%%%%%%%%%%%%%%%%%%%%%%%%%%%%%%%%%%%%%%%%%%%%%%%%%%%%%%
\section{Theory}
\label{sec:theory}

Consider a diatomic molecular system consisting of one electron
and two nuclei of arbitrary charges $Z_1$ and $Z_2$ placed at a
fixed distance $R$ along a line defining the $z$ axis; let $r_1$
denote the distance of the electron from nucleus $1$ and $r_{2}$
from nucleus $2$.
To simplify we neglect
any nuclei finite mass effect.

In  prolate spheroidal coordinates, defined by
\begin{equation}
    \xi \equiv \frac{r_{1}+r_{2}}{R} \,;  \hspace{1cm}
    \eta \equiv \frac{r_{1}-r_{2}}{R} \,; \hspace{1cm}
    \phi \equiv \arctan \left( \frac{y}{x} \right) \,
    \label{eq:coordprol}
\end{equation}
where $1\leq \xi < \infty$, $-1\leq \eta\leq 1$ and $0\leq \phi
\leq 2\pi$, the Schr\"odinger equation for the electron reads
\begin{eqnarray}
\bigg\{
-\frac{2}{R^{2}( \xi ^{2}-\eta ^{2} )}
 \bigg [
\frac{\partial }{\partial \xi } ( \xi ^{2}-1 )
                \frac{\partial }{\partial \xi }  +
\frac{\partial }{\partial \eta } ( 1-\eta ^{2} )
               \frac{\partial }{\partial \eta } + \nonumber \\
+ \frac{\xi ^{2}-\eta ^{2}}{( \xi ^{2}-1 )( 1-\eta ^{2} )}
\frac{\partial^2 }{\partial \phi ^2}
 \bigg ] +V(\eta,\xi)
\bigg \} \, \psi (\xi,\eta,\phi )=E \,  \psi \, (\xi,\eta,\phi ) \, ,
\label{eq:H2+schro2}
\end{eqnarray}
with the electron-nuclei potential given by
\begin{equation}
V(\xi,\eta)=-\frac{Z_{1}}{r_{1}}-\frac{Z_{2}}{r_{2}} \, =
-\frac{2}{R} \, \frac{(Z_{1}+Z_{2})\xi-(Z_{1}-Z_{2})\eta }
{(\xi ^{2}-\eta^{2})} \, .
\end{equation}
In the fixed--nuclei approximation, the internuclear
distance $R$ enters as a parameter, and the nuclei repulsive
potential energy $1/R$  may be simply added. Equation
(\ref{eq:H2+schro2}) is separable in these coordinates, meaning
that the solution is expressed as a product of three  functions
\begin{equation}
    \psi(\xi ,\eta ,\phi )=U(\xi )\Lambda(\eta )\Phi (\phi ) \, .
    \label{eq:H2+12}
\end{equation}
%
%\subsection{H$_2^+$ molecule in prolates coordinates}
%
The azimuthal function $\Phi$ is easily separated, and  must
fulfill the equation
\begin{equation}
    \frac{\mathrm{d}^{2} \Phi}{\mathrm{d} \phi^{2}}+m^{2}\Phi=0 \, ,
    \label{eq:phi22}
\end{equation}
whose solutions are
\begin{equation}
    \Phi (\phi )=\frac{1}{\sqrt{2\pi}} \, e^{im\phi} \, ,
    \label{eq:phi2}
\end{equation}
with $m=0,\pm 1,\pm 2, \pm 3 ,\cdots $.
Because of the axial symmetry of the potential, $m$ is a good quantum number.

Upon elimination of the azimuthal dependence, and defining
$p^{2}=-\frac{R^{2}E}{2}$, $a_{1}=R(Z_{1}-Z_{2})$ and
$a_{2}=R(Z_{1}+Z_{2})$, the ensuing equation reads
\begin{eqnarray}
\bigg \{
&& \frac{\partial }{\partial \xi }
  \left [( \xi^2-1 )\frac{\partial }{\partial \xi } \right ]+
 a_2 \xi -p^2 \xi^{2} - \frac{m^{2}}{\xi^{2}-1} +
 \\
&+&  \frac{\partial }{\partial \eta }
  \left [ ( 1-\eta^2  )\frac{\partial }{\partial \eta } \right ] -
  a_1\eta + p^2 \eta^{2}- \frac{m^2}{1-\eta^2}
\bigg \} \, U(\xi)\Lambda(\eta) = 0 \, .\nonumber
\label{eq:separation}
\end{eqnarray}
and is also separable. Denoting the separation constant as $A$,
one obtains a system of two non--trivial ordinary differential
equations, a ``radial" equation for $U(\xi)$ and an ``angular"
equation for $\Lambda(\eta)$,
\begin{subequations}
\begin{eqnarray}
%\begin{equation}
    &&\left [ \frac{\partial }{\partial \xi }\left [\left ( \xi
    ^{2}-1\right )\frac{\partial }{\partial \xi }\right ]+ a_{2}\xi
    -p^{2}\xi ^{2}-\frac{m^{2}}{\xi ^{2}-1}+ A \right ]U(\xi )=0
    \, ,
    \label{eq:xiseparada} \\
%\end{equation}
%\begin{equation}
    &&\left [ \frac{\partial }{\partial \eta  }\left [\left (1- \eta
    ^{2}\right )\frac{\partial }{\partial \eta  }\right ]- a_{1}\eta
    +p^{2}\eta  ^{2}-\frac{m^{2}}{1-\eta  ^{2}}- A \right ]
     \Lambda(\eta)=0 \, ,
    \label{eq:etaseparada}
\end{eqnarray}
\end{subequations}
which are coupled through both the scaled energy $p$ and the
coupling constant $A$. States with different $m$ values
are not coupled, so that they can be considered independently.

In this work, we propose two different methods using a spectral
approach based on GSF in prolate spheroidal coordinates. 
In the first -- named hereafter ``iterative $1d$ method'' -- we
solve, alternately, the one--dimensional radial equation 
(\ref{eq:xiseparada}), assuming a given separation constant $A$,
and solving an eigenvalue equation for the scaled energy $p$. 
Then, we use this energy as a fixed value in the one--dimensional 
angular equation (\ref{eq:etaseparada}), obtaining a new separation 
constant $A$. 
The process is repeated until convergence is achieved.
In this iterative procedure, both
equations are solved by using adequate GSF basis sets and are
converted into eigenvalue problems. The main advantage of our GSF
approach resides in the fact that the principal part of these two
equations (in particular, the derivatives) are already dealt with
by the basis functions; as a consequence,  derivative calculations
are not required at every iteration step. In the second method, we
construct a basis set composed of products of the angular and
radial GSF. This two--dimensional basis is used to represent the
Hamiltonian, which is diagonalized in order to solve the whole
Schr\"odinger equation (\ref{eq:H2+schro2}). In this way, we
obtain the eigenvalues (energies) and eigenvectors (solutions) of
many states at the same time. This method, here referred to as the
``direct $2d$ method", while possessing the same advantages
related to GSF is computationally even more efficient.

%%%%%%%%%%%%%%%%%%%%%%%%%%%%%%%%
%%%%%%%%%%%%%%%%%%%%%%%%%%%%%%%%
\subsection{GSF: iterative $1d$ method}
\label{subsec:iterative}

\subsubsection{Angular equation} %\hfill\\

\begin{figure}[h!]
\centering
\includegraphics[width=0.7\textwidth]{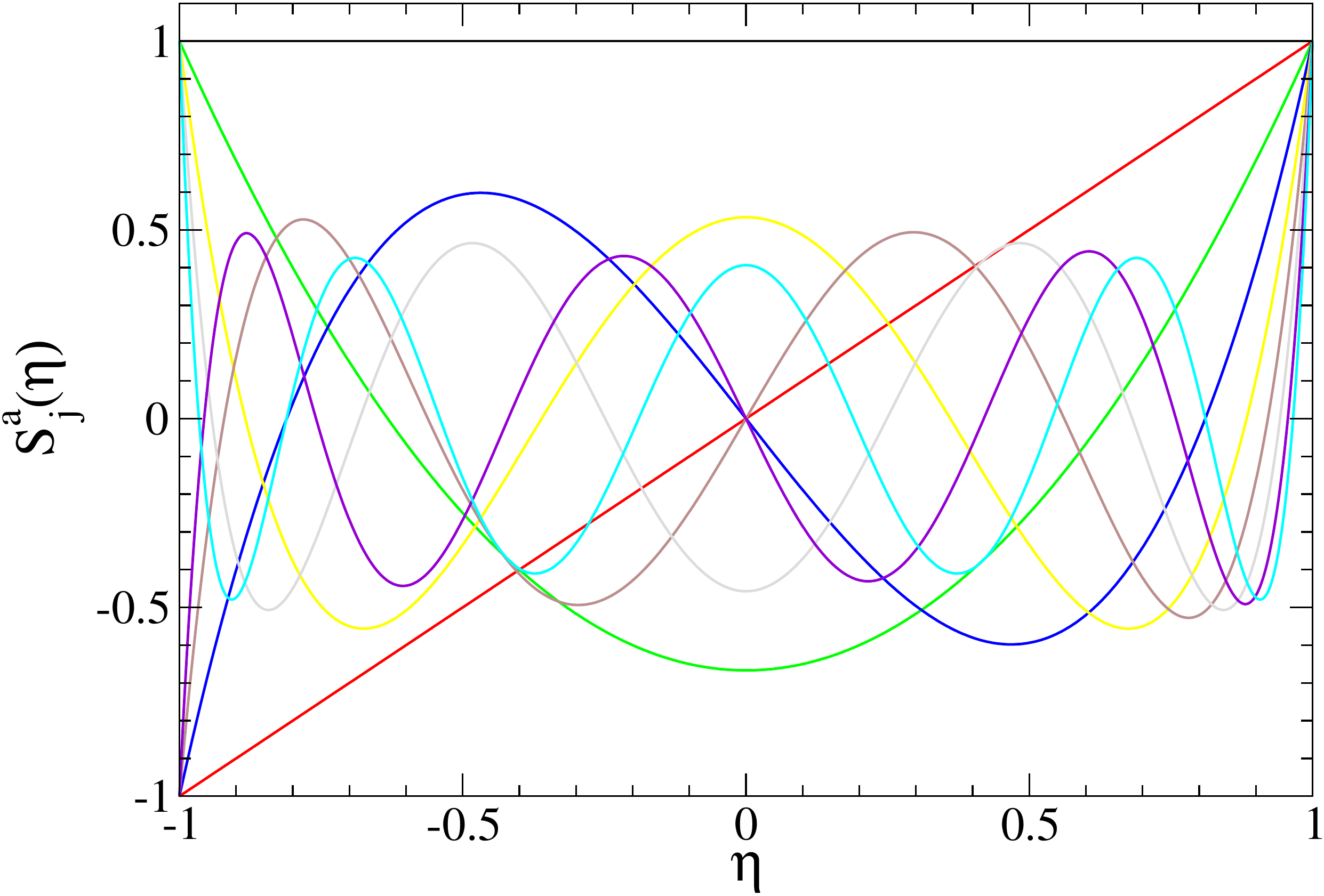}
\caption{First 9 angular Sturmian basis elements $S^a_j(\eta)$ for
$m=0$.} \label{fig:etabasis}
\end{figure}

We search the solution of Eq. (\ref{eq:etaseparada}), for a given
$m$, as an expansion in Sturmian functions
\begin{equation}
\Lambda(\eta )=\sum_{j} \, c_{j} \, S^a_{j}(\eta) \, ,
\label{eq:expansionlambda}
\end{equation}
the angular basis set being generated by solving the Sturmian
equation
\begin{equation}
    \left[\frac{\partial }{\partial \eta  }\left [\left (1- \eta
    ^{2}\right )\frac{\partial }{\partial \eta  }\right] -\frac{m^{2}}{1-\eta  ^{2}}\right ] \,S^a_{j}(\eta)
    =-\beta _{j} \, S^a_{j}(\eta) \, ,
\label{eq:steta}
\end{equation}
with boundary conditions $S^a_{j}(1)=1$ and $S^a_{j}(-1)=(-1)^j$ for
$m=0$ and $S^a_{j}(1)=S^a_{j}(-1)=0$ for $m\neq 0$. The solutions are
 actually the well known associated Legendre
polynomials \cite{Edmonds}, $S^a_j(\eta)=P_j^m(\eta)$,
and correspond to eigenvalues $\beta _{j}=j(j+1)$. Figure
\ref{fig:etabasis} shows the first 9 elements $S^a_j(\eta)$ for
$m=0$.

With expansion (\ref{eq:expansionlambda}) and making use of Eq.
(\ref{eq:steta}), the angular equation (\ref{eq:etaseparada})
becomes
\begin{eqnarray}
\sum_j c_j \,
\left [ -\beta_j - a_{1}\eta
    +p^{2}\eta  ^{2}  \right ] \, S^a_j(\eta ) =
 A \, \sum_j c_j \,  S^a_j(\eta ) .
\label{eq:etaP}
\end{eqnarray}
Multiplying from the left by $S^a_i(\eta)$ and integrating over the
angular domain $[-1,1]$, we obtain a generalized eigenvalues
equation
\begin{eqnarray}
\mathbf{M} \, \mathbf{c}  = A \, \mathbf{B} \, \mathbf{c} \,
.\label{eq:etaMatriz}
\end{eqnarray}
 The matrices involve the elements
\begin{eqnarray}
[\mathbf{{\cal M}^k}]_{ij} &=& \int_{-1}^{1} S^a_i(\eta) \, \eta^k
\, S^a_j(\eta) \, d\eta  \label{eq:Mk}
\end{eqnarray}
which can be evaluated analytically using known properties of the
Legendre polynomials  \cite{Edmonds}. Those of interest here are
given by
\begin{subequations}
\label{Melements}
\begin{eqnarray}
 \left[\mathbf{{\cal
M}^0}\right]_{ij}
&=&  \, \frac{2}{2i+1} \frac{(i+m)!}{(i-m)!} \delta_{ij} \,
\label{eq:M1}
\\
 \left[\mathbf{{\cal
M}^1}\right]_{ij}
&=&  \frac{2}{2i+1} \frac{(i+m)!}{(i-m)!} \frac{1}{2j+1} \left[
(j-m+1) \, \delta _{i,j+1} + (j+m) \, \delta_{i,j-1} \right] \,
\label{eq:M2}
\\
 \left[\mathbf{{\cal
M}^2}\right]_{ij}
&=&  \frac{2}{2i+1} \frac{(i+m)!}{(i-m)!} \frac{1}{2j+1} \bigg[
    \frac{(j+1-m)(j+2-m)}{2j+3} \, \delta _{i,j+2}
    \nonumber\\
    &+&
\left (
    \frac{(j+1-m)(j+1+m)}{2j+3}+\frac{(j+m)(j-m)}{2j-1}
\right ) \, \delta _{i,j} \nonumber \\   &+&
\frac{(j-1+m)(j+m)}{2j-1} \, \delta _{i,j-2} \bigg] \, ,
\label{eq:M3}
\end{eqnarray}
\end{subequations}
and are calculated only once, at the first iteration.
The elements of the matrices $\mathbf{M}$ and $\mathbf{B}$ are
given by
\begin{subequations}
\begin{eqnarray}
\left[\mathbf{M}\right]_{ij} &=& -j(j+1) \, [\mathbf{{\cal
M}^0}]_{ij} - a_1 \, [\mathbf{{\cal M}^1}]_{ij} + p^2 \,
[\mathbf{{\cal M}^2}]_{ij}
\\
\left[\mathbf{B}\right]_{ij} &=& [\mathbf{{\cal M}^0}]_{ij} \, .
\label{eq:B}
\end{eqnarray}
\end{subequations}

Assuming a given energy value $p^2$, the angular part reduces to
solving the generalized eigenvalues problem (\ref{eq:etaMatriz}),
\emph{i.e.}, finding the eigenvalue $A$ (the separation constant)
and the eigenvector $\mathbf{c}$ (the  coefficients of expansion
(\ref{eq:expansionlambda})). At each iteration, the matrix
$\mathbf{M}$ is easily recalculated with the new energy value $p$.

%%%%%%%%%%%%%%%%%%%%%%%%%%%%%%%%
%%%%%%%%%%%%%%%%%%%%%%%%%%%%%%%%
\subsubsection{Radial equation}  %\hfill\\
\label{subsubsec:radial}

Once the $A$ eigenvalue is obtained from the angular equation, the
scaled energy $p^2$ is to be found from solving the radial
equation (\ref{eq:xiseparada}). Setting $U(\xi) = (\xi^2-1)^{\vert
m \vert/2} f(\xi)$ removes the singular term $m^2/(\xi^2-1)$ from
the differential equation.
A first  boundary condition is
\begin{equation}
\lim_{\xi \to \infty} \, f(\xi ) = e^{-p \xi}  \, .
\label{eq:asympxiinf}
\end{equation}
We can set a second boundary condition at the other end, when the electron 
is exactly in the center of the molecular system ($\xi = 1$). 
We have to distinguish two cases.
When $m = 0$
\begin{equation} \lim_{\xi \to 1} \, f(\xi ) =
\xi^{-\frac{A}{2}} \, e^{\frac{p^2}{4} \xi^{2} -
    \frac{a_2}{2}\xi} \, .
\label{eq:asympxi1}
\end{equation}
because  the radial equation (\ref{eq:xiseparada}) reduces to
\begin{eqnarray}
\frac{d f(\xi)}{d \xi} &=& \left( \frac{p^2}{2}\xi - \frac{a_2}{2}
- \frac{A}{2 \xi} \right) f(\xi ) \, .
\end{eqnarray}
For $m\neq 0$, the function $U(\xi)$ will vanish at $\xi=1$ as long as $f(\xi)$ 
does not present any singularity at that value.

\begin{figure}[h]
\centering
\includegraphics[width=0.7\textwidth]{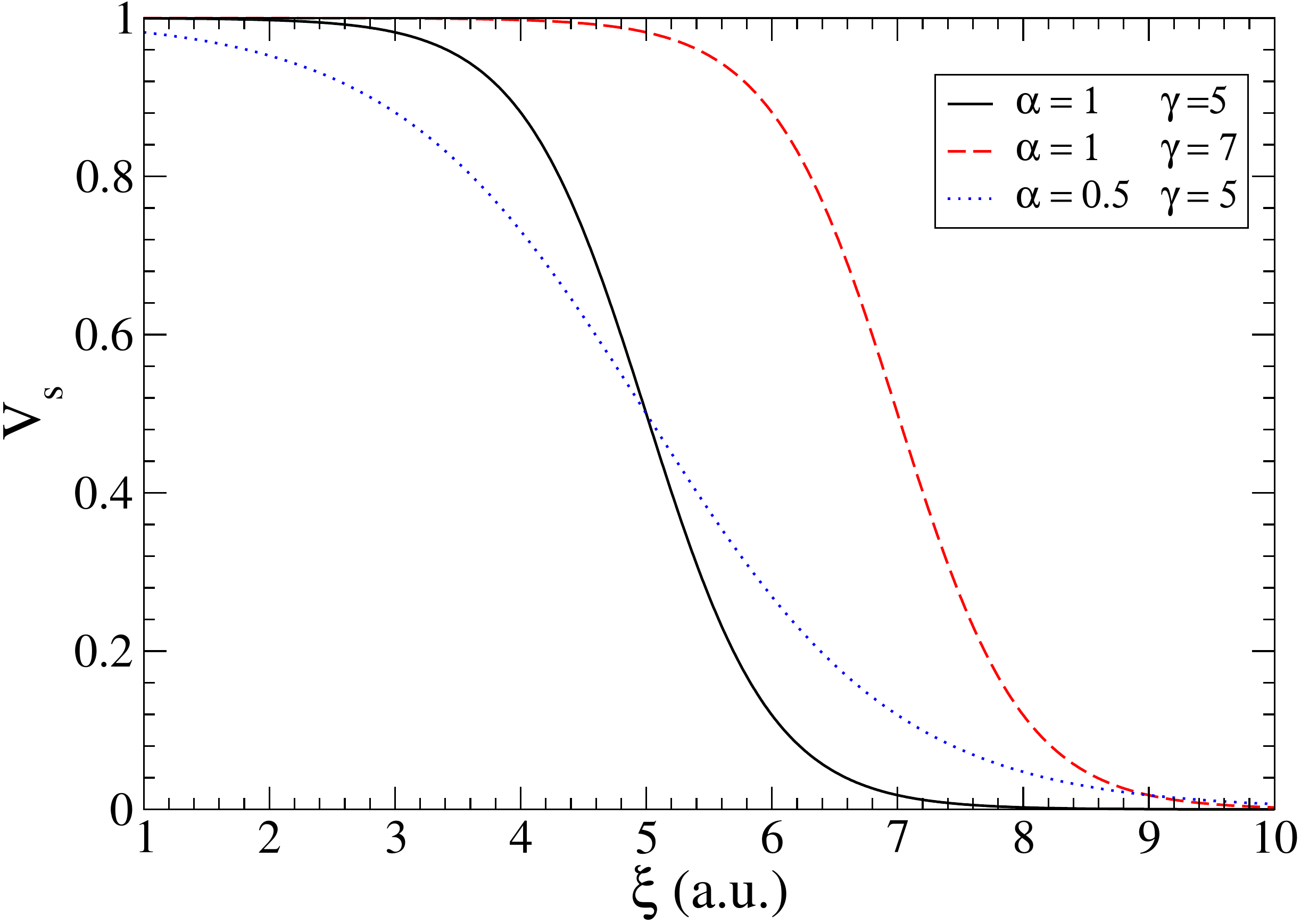}
\caption{Generating potential $V_{\mathrm{s}}$, used to generate
the radial GSF ${\cal S}^r_j(\xi)$.} \label{fig:Vs}
\end{figure}

Similarly to the angular part, we propose an expansion
\begin{equation}
U(\xi) = (\xi^2-1)^{\vert m \vert/2} \sum_j d_j {\cal S}^r_j(\xi) \,
, \label{eq:expansionxi}
\end{equation}
on a basis of Generalized Sturmian Functions ${\cal S}^r_j(\xi)$
generated by the Sturmian equation
\begin{equation}
    \left [ \frac{\partial }{\partial \xi }\left [\left ( \xi
    ^{2}-1\right )\frac{\partial }{\partial \xi }\right ]
    +   2\xi \vert m \vert \frac{\partial }{\partial \xi}
+    a_{2} \, \xi -p_{s}^{2} \, \xi ^{2} \right ] \, 
{\cal S}^r_{j}(\xi ) = \alpha_{j} \, V_{\mathrm{s}}(\xi)  \, 
{\cal S}^r_{j}(\xi ) \, , \label{eq:sturxi}
\end{equation}
with eigenvalues $\alpha_{j}$. In Eq. (\ref{eq:sturxi}),
$p_s^{2}=-\frac{R^{2}E_s}{2}$ is a parameter that can be set
freely. However, since the expansion over the GSF basis is meant to 
represent a physical radial function, it is convenient and numerically efficient 
to choose $E_s$ according to the physics one wishes to describe. 
When dealing with a continuum state of energy $E>0$, taking $E_s=E$ is a natural 
choice.  
In order to represent a specific bound state with an a priori unknown energy 
value, taking $E_s<0$ close to a guess of the sought after energy turns out to 
be a good choice. 
In both continuum and bound cases, an appropriate choice of $E_s$ will impose 
an adequate energy behavior onto the GSF functions, ultimately making the basis 
more efficient from a convergence point of view. 
$V_s$, known as generating 
potential, must be a short range potential so that the basis elements
${\cal S}^r_{j}(\xi )$ have an asymptotic behavior similar to 
(\ref{eq:asympxiinf}), that is to say an exponential decay with energy 
$E_s$ (taking $E_s$ close to the correct sought after value $E$ is then a 
natural choice).  
Moreover, since we wish
${\cal S}^r_{j}(\xi )$ to possess also the same $\xi \to 1$ behavior
as the sought after solution $U(\xi)$, the generating potential
must obey the relation
\begin{equation}
\lim_{ \xi \rightarrow 1 } \,\, \alpha_j \, V_s(\xi)  = -A + p^2
-p_s^2\, .
\end{equation}
It turns out that is convenient to choose a function nearly
constant at $\xi = 1$, in order to stabilize the iterations. In
the present work, the generating potential is chosen to be
\begin{equation}
V_{\mathrm{s}}=\frac{1}{2}
\left[ 1 - \tanh(\delta \, (\xi-\gamma) ) \right] \, ,
    \label{eq:potVshort}
\end{equation}
where the parameters $\delta$ and $\gamma $  determine the shape
of the potential as illustrated by Figure \ref{fig:Vs}.
For a given value of $\delta$, a larger parameter $\gamma$ extends the
range of the potential (for $\delta=1$, $\gamma$ approximately
represents the range). On the other hand, for a fixed value of
$\gamma$ (solid and dotted curves), higher $\delta$ parameters
correspond to steeper potentials.
As explained in the GSF references \cite{Mitnik:11,Gasaneo:13}, 
the generating potential is crucial for the continuum functions. 
For bound type solutions, on the other hand, the choice of $V_s$ 
is not so important (it does not affect noticeably the convergence of the method) 
but helps for example in regulating the radial domain covered by the GSF.  
We have not performed an exhaustive optimization of the potential parameters, 
but we found, roughly, that changing these values by an order of magnitude
affects the final bound state energy values only beyond the sixth significant 
figure.
As a rule of thumb, our numerical investigation established that 
the values $\delta \approx 1$ and $\gamma \approx 5$ are a suitable choice 
for the potential parameters in the case of the ground state.
For excited states with principal quantum number $n$, the potential 
range should be incremented roughly by a factor 
$\frac{\Delta \gamma}{\Delta n}  \approx 2$.
Also, for varying internuclear distances $R$, it is convenient to 
scale the potential range by a factor $\frac{2}{R}$.

\begin{figure}[h]
\centering
\includegraphics[width=0.7\textwidth]{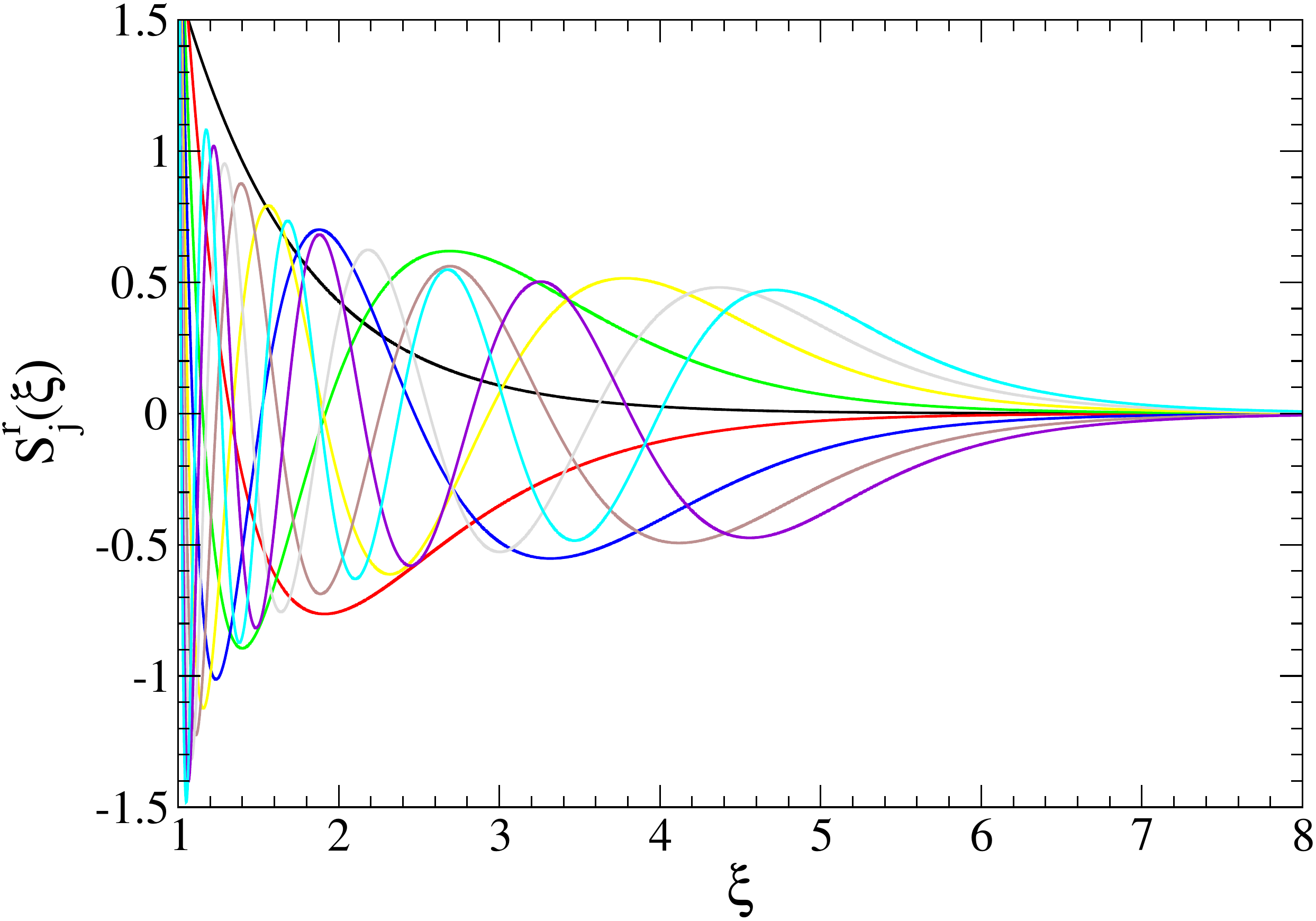}
\caption{First 9 radial basis elements ${\cal S}^r_j(\xi)$ for
$m=0$.} \label{fig:xibasis}
\end{figure}

At $\xi \to \infty$ we could impose on ${\cal S}^r_j(\xi)$ the
 boundary condition  (\ref{eq:asympxiinf}), but
requiring simply the basis function to vanish at infinity was found to be
sufficient. On the other hand, imposing on each element condition
(\ref{eq:asympxi1}) at $\xi \to 1$ results to be crucial when
$m=0$. 
We generate the Sturmian functions by solving the radial equation 
(\ref{eq:sturxi}) with a finite difference method. 
In Ref \cite{Mitnik:11} the reader can find a 
detailed description of the numerical procedures used for the solution 
of the differential equation, which in turn, are based on the radial 
methods for the solution of the Schr\"odinger equation described in 
W. Johnson's book \cite{Johnson:07}. 
Briefly, the solution integration consists of a
predictor--corrector Adams--Moulton method.
It uses a seven--point scheme, which (together with the 
interpolation procedure) achieves a high order of accuracy 
(of about $(\Delta x)^8$). 
The original GSF code was developed primarily for Coulomb--type solutions and 
for high principal quantum numbers. 
Since these functions oscillate rapidly close to the nucleus and decay 
exponentially far away, one may use a logarithmic grid generating 
a fine mesh near the origin and a coarse mesh for large distances. 
With this approach, very accurate results can be obtained by using only 
a few points (about 500) in the numerical grid.
Since in the present investigation we are interested in the first 
eigenfunctions we can relax the numerical sophistication and complexity, 
and use a low--order Numerov approximation for the propagation, in a linear mesh. 
Of course, this replacement would require a large number of mesh points 
(about $10^4$), but this is not a serious problem in a one--dimensional 
calculation. 
The numerical quadratures are evaluated using a trapezoidal rule with endpoint 
corrections developed by Johnson \cite{Johnson:07}.
The first 9 basis elements  for $m=0$, generated with $\delta=1.1$ and
$\gamma=5$, are shown in Figure \ref{fig:xibasis}. As $j$
increases, these functions display an increasing number of nodes.
Featuring one of the main GSF properties, all elements behave
asymptotically in a unique manner, here in the same exponential
manner $e^{-p_s \xi}$ as $\xi \to \infty$.

With  expansion (\ref{eq:expansionxi}), and making use of
(\ref{eq:sturxi}), the radial equation (\ref{eq:xiseparada}) takes
the form
\begin{equation}
\sum_j d_j \left [ \alpha_j \, V_{\mathrm{s}}(\xi)+ A +m^2 + \vert
m \vert\right ] \, {\cal S}^r_j(\xi ) = \sum_j d_j (p^2 - p_s^2) \,
\xi^2 \, {\cal S}^r_j(\xi) \, .\label{eq:xiespansion}
\end{equation}
Multiplying from the left by ${\cal S}^r_i$ and integrating over the
domain $[1,\infty[$, we obtain another generalized eigenvalues
equation
\begin{equation}
\mathbf{N} \, \mathbf{d} = \lambda \, \mathbf{C} \, \mathbf{d}
\label{eq:autovaloresradial}
\end{equation}
where the eigenvalues are $\lambda = p^2 - p_s^2$, and thus the
corresponding energies through $p^2=-R^2  E/2$. Let us define the
elements
\begin{subequations}
\begin{eqnarray}
\left[\mathbf{{\cal N}^k}\right]_{ij} &=& \int_{1}^{\infty} 
{\cal S}^r_i(\xi)
\, \xi^k \, {\cal S}^r_j(\xi) \, d\xi  \label{eq:Nk} \\
\left[\mathbf{G}\right]_{ij} &=&  \int_1^\infty \, {\cal S}^r_i(\xi)
\, V_{\mathrm{s}}(\xi)  \, {\cal S}^r_j(\xi) \, d\xi \, ,
\label{eq:G}
\end{eqnarray}
\end{subequations}
that are calculated, numerically, only once. The
matrices $\mathbf{N}$ and $\mathbf{C}$ have for elements
\begin{subequations}
\begin{eqnarray}
\left[\mathbf{N}\right]_{ij} &=& \, (A+m^2 + \vert m \vert) \,
[\mathbf{{\cal N}^0}]_{ij} \,
+\alpha_j \, [\mathbf{G}]_{ij} \label{Nmatrix}\\
\left[\mathbf{C}\right]_{ij} &=& [\mathbf{{\cal N}^2}]_{ij} \, .
\label{Cmatrix}
\end{eqnarray}
\end{subequations}
Here $A$ is a fixed parameter obtained from the previous step,
 when solving the angular part. The solutions of (\ref{eq:autovaloresradial})
provide both the eigenvalues $\lambda$ and the eigenvectors made
of the coefficients $d_j$ of the radial expansion
(\ref{eq:expansionxi}).

This iterative method has a significant advantage. The Hamiltonian
is separated into two coupled equations, and both of them are
one--dimensional reducing significantly the computational cost.
Moreover, the use of expansions on GSF basis greatly simplifies
the task since each basis element already solves a substantial
part of the equations, in particular the differential operators.
As a consequence,  it is not necessary to solve numerically the
differential equations at each step. Computationally, one only
solves -- iteratively -- two generalized eigenvalue problems. There
is, however, a drawback in this methodology: each molecular state
requires a new basis set. This means that, from all the
eigenvalues $A$ and $p$ resulting from the calculations, we must
select only those corresponding to the eigenvectors having the
right number of nodes. For each one of the molecular states, a
different iteration procedure is thus needed. This difficulty is
avoided in the alternative method presented hereafter.

%%%%%%%%%%%%%%%%%%%%%%%%%%%%%%%%
%%%%%%%%%%%%%%%%%%%%%%%%%%%%%%%%
\subsection{GSF: direct $2d$ method}
\label{subsec:direct}

We propose now a method in which equation (\ref{eq:H2+schro2}) is
solved directly. As before, we first remove the azimuthal part and
write
\begin{equation}
\psi (\xi,\eta,\phi) = \Psi(\xi,\eta) \Phi(\phi)
\end{equation}
with $\Psi(\xi,\eta)$ solution of the two-dimensional equation
\begin{eqnarray}
&\bigg \{ &
 \frac{\partial }{\partial \xi }
  \left [( \xi^2-1 )\frac{\partial }{\partial \xi } \right ]+
 a_2\xi - p^2 \, \xi^{2} -\frac{m^{2}}{\xi ^{2}-1} \nonumber \\
+  && \frac{\partial }{\partial \eta }
  \left [ ( 1-\eta^2  )\frac{\partial }{\partial \eta } \right ]
  - a_1 \eta + p^2 \, \eta^{2} -\frac{m^{2}}{1-\eta^{2}}
\bigg \} \, \Psi(\xi,\eta) = 0 \, . \label{SE2d}
\end{eqnarray}
This time we propose to expand the solution $\Psi (\xi,\eta)$ over
a two--dimensional basis $S_{ij}(\xi,\eta)$
\begin{eqnarray}
\psi (\xi,\eta) = \sum_{ij} \, a_{ij} \, {\rm S}_{ij}(\xi,\eta) \,
= (\xi^2-1)^{\vert m \vert/2} \, \sum_{ij} \, a_{ij} \, 
{\cal S}^r_i(\xi) \, S^a_j(\eta) \, \label{expansion2d}
\end{eqnarray}
where the one--dimensional Sturmian functions are obtained with
the same methodology described above, \emph{i.e.}, from equations
 (\ref{eq:steta}) and (\ref{eq:sturxi}).

Upon substitution of expansion (\ref{expansion2d}), the two--dimensional
equation (\ref{SE2d}) becomes
\begin{eqnarray}
&&\sum_{ij} \, a_{ij} \bigg \{ \alpha_i \, V_{\mathrm{s}}(\xi) +
m^2 + \vert m \vert + p_s^2 \, \xi^2 - a_1 \eta - \beta_j \bigg \}
\, {\rm S}_{ij}(\xi,\eta) \nonumber \\ &=&  \sum_{ij} \, a_{ij}
\, p^2  \, (\xi^2 - \eta^2) \, {\rm S}_{ij}(\xi,\eta)
 \, .
\label{eq:H2dstur}
\end{eqnarray}
A matrix system is constructed by multiplying from the left by a
basis element ${\rm S}_{i'j'}(\xi,\eta)$ and integrating over both
$\xi$ and $\eta$ variables (note here the absence of the volume
element $\xi^2-\eta^2$ in spheroidal prolate coordinates). We
obtain a generalized eigenvalues problem
\begin{equation}
\mathbf{P} \, \mathbf{a} = \mathbf{\lambda} \, \mathbf{D} \,
\mathbf{a} \, ,
\end{equation}
in which the matrices $\mathbf{P}$ and $\mathbf{D}$ are given by
\begin{subequations}
\begin{eqnarray}
\left[\mathbf{P}\right]_{i'j',ij} &=&
 \alpha_i \,
[\mathbf{G}]_{ii'}  \left[\mathbf{{\cal M}^0}\right]_{j'j} + p_s^2
\, \left[\mathbf{{\cal N}^2}\right]_{i'i}
[\mathbf{{\cal M}^0}]_{j'j}  \nonumber \\
& & - a_1 \left[\mathbf{{\cal N}^0}\right]_{i'i} [\mathbf{{\cal
M}^1}]_{j'j} +(m^2 + \vert m \vert - \beta_j) \,
\left[\mathbf{{\cal N}^0}\right]_{i'i} [\mathbf{{\cal M}^0}]_{j'j}
\label{Pelements}
\\
\left[\mathbf{D}\right]_{i'j',ij} &=&
\left[\mathbf{{\cal N}^2}\right]_{i'i}  [\mathbf{{\cal M}^0}]_{j'j}
- \left[\mathbf{{\cal N}^0}\right]_{i'i}  [\mathbf{{\cal M}^2}]_{j'j} \, .
\label{Delements}
\end{eqnarray}
\end{subequations}
We solve this eigenvalue problem, obtaining a solution matrix
$\mathbf{a}$; each column consists of the coefficients
vector ${\vec a^n}$, which expands that solution corresponding to
the molecular state having eigenenergy $\lambda_n = p_n^2$. To be
more specific, if the basis size is $N$, we have
\begin{equation} \mathbf{a} =
\begin{pmatrix}
a^1_{11} & a^2_{11} & a^3_{11} & \ldots & a^N_{11} \\
a^1_{21} & a^2_{21} & a^3_{21} & \ldots & a^N_{21} \\
\ldots & \ldots & \ldots & \ldots & \ldots \\
a^1_{12} & a^2_{12} & a^3_{12} & \ldots & a^N_{12} \\
a^1_{22} & a^2_{22} & a^3_{22} & \ldots & a^N_{22} \\
\ldots & \ldots & \ldots & \ldots & \ldots \\
a^1_{ij} & a^2_{ij} & a^3_{ij} & \ldots & a^N_{ij} \\
\ldots & \ldots & \ldots & \ldots & \ldots
\end{pmatrix}
\, \, \,
\mathbf{\lambda} =
\begin{pmatrix}
p_1^2 \\
p_2^2 \\
\ldots \\
p_N^2
\end{pmatrix}
\end{equation}

This direct methodology avoids iterations. Moreover, it allows us
to obtain the solutions for many molecular states simultaneously.
Since the matrices are two--dimensional, at first sight the method
seems computationally costly. However, all integrations leading to
the matrix elements of $\mathbf{P}$ and $\mathbf{D}$ are separable
and reduce to products of one--dimensional integrals as given by
(\ref{Pelements}) and (\ref{Delements}).

%%%%%%%%%%%%%%%%%%%%%%%%%%%%%%%%%%%%%%%%%%%%%%%%%%%%%%%%%%%%%%%%%%%%%%
\section{RESULTS}
\label{sec:results}

We  present now the results of our calculations and make a
comparison with the data provided in the literature. We start by
applying the GSF iterative $1d$ method for both the ground and
some $m=0$ excited states of the hydrogen molecular ion
H$_{2}^{+}$ for which $Z_{1}=Z_{2}=1$ and thus $a_1=0$. Next we
consider asymmetric (heteronuclear) molecular ions with $Z_{1}\neq
Z_{2}$. Finally, for H$_{2}^{+}$, we will show how the GSF direct
$2d$ method yields the ground and several excited states in a
single run.

%%%%%%%%%%%%%%%%%%%%%%%%%%%%%%%%%%%%%%%%%%%%%%%%%%%%%%%%%%%%%%%%%%%%%%
\subsection{Iterative $1d$ method for the ground state of H$_2^+$}
\label{subsec:iterativeH2}

The best values of the energy $E$ for the ground state
$1\sigma_{g}$, and the corresponding separation constant $A$ from
the work of Scott {\it et al.} \cite{Scott} are used here as a
benchmark to analyze the convergence issues of our Sturmian
method. We assume here an internuclear distance $R=2$ a.u., thus
fixing the values of $a_1$ and $a_2$.

%%%%%%%%%%%%%%%%%%%%%%%%%%%%%%%%%%%%%%%%%%%%%%%%%%%%%%%%%%%%%%%%%%
\subsubsection{Angular equation} %\hfill\\

In order to solve the angular equation (\ref{eq:etaseparada}), an
initial value for the energy $E$ (more precisely, $p^2=2.2052684$)
is chosen. The matrices of the generalized eigenvalue problem
(\ref{eq:etaMatriz}) are easily constructed as they are all
analytical. The only numerical aspect to analyze is the
convergence of the results with respect to the basis size. Since
the ground state is an even function in the $\xi$ coordinate, only
even elements $S^a_j(\eta)$ are included in the expansion. As shown
through Table \ref{table:convergenciaA}, convergence towards the
benchmark result $A$ of Ref. \cite{Scott} is reached with just 4
elements.

\begin{table}[h]
\centering
\begin{tabular}{| c | c |}
\hline
\rule{0pt}{1.1\normalbaselineskip} 
  Number of basis elements  &  \textbf{$A$}\\
\hline
\textbf{$1$} & 0.7350895\\
\hline
\textbf{$2$} & 0.8115139\\
\hline
\textbf{$3$} & 0.8117295\\
\hline
\textbf{$4$} & 0.8117296\\
\hline
Reference \cite{Scott} & 0.8117296\\
\hline
\end{tabular}
\caption{Convergence of the eigenvalue $A$ in Eq.
(\ref{eq:etaseparada}) for fixed energy $E=1.10264$, as a function
of the basis size.} \label{table:convergenciaA}
\end{table}

\begin{figure}[h!]
\centering
\includegraphics[width=0.7\textwidth]{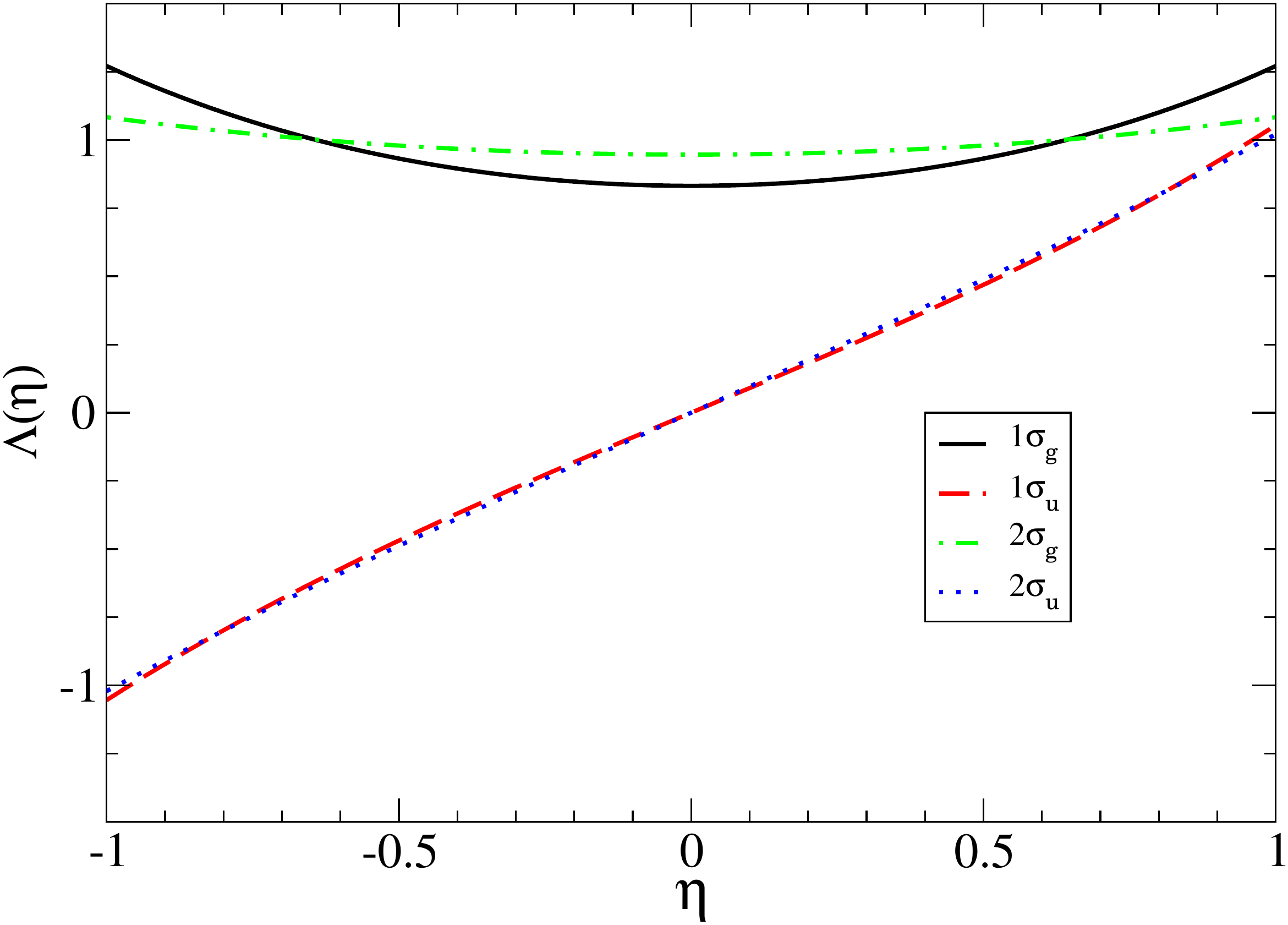}
 \caption{The angular $\Lambda(\eta)$
solutions for the four lower energy states of H$_{2}^{+}$.}
\label{fig:4estadoseta}
\end{figure}

Having solved the matrix equation, the eigenvectors give the
coefficients $c_j$ that allow us to construct the ground state
angular solution (\ref{eq:expansionlambda}) which is shown in
Figure \ref{fig:4estadoseta}. The excited states will be discussed
in the next section.

\newpage
%%%%%%%%%%%%%%%%%%%%%%%%%%%%%%%%%%%%%%%%%%%%%%%%%%%%%%%%%%%%%%%%%%
\subsubsection{Radial equation} %\hfill\\

Once the angular equation is solved, we turn to the radial
equation. In contrast to the angular part, here the approach is
completely numerical. On one hand we have to generate the basis
set ${\cal S}^r_i(\xi)$ and, on the other,  the matrix elements of
the corresponding eigenvalue problem (\ref{eq:autovaloresradial})
must be calculated numerically.

The basis elements are generated by solving the Sturmian equation
(\ref{eq:sturxi}) with a numerical method described previously
\cite{Mitnik:11}. It is based on a predictor--corrector algorithm,
propagating the solution from the origin to some defined matching
point (this is the outgoing solution), and from an effective
infinite towards this point (the inward solution). The inward
function is normalized, in such a way that both solutions coincide
at the matching point. If the derivatives disagree at this point,
the eigenvalue is adjusted and the procedure starts again, until
convergence. This algorithm, allows one to produce very accurate
solutions for atomic systems, even with a reasonably small ($\sim
500$) points in the numerical grid \cite{Mitnik:11,Gasaneo:13}.
However, we noticed that it was hard to obtain the radial
solutions of Eq. (\ref{eq:xiseparada}), even when a large number
of points was included in the numerical grid. In fact, to solve
this equation appropriately, the crucial aspect resides in the
fulfillment of the boundary conditions (\ref{eq:asympxi1}) at the
origin. We endorsed this conclusion, trying to solve the radial
equation with other methods, and using different mathematical
softwares, obtaining very different results for different
numerical grids. We even tried to solve the equation fixing the
energy value to $E=-1.10264$ a.u. \cite{Scott}, but the converged
solutions yielded eigenvalues $A$ too far from the correct value.

We also tried to use standard diagonalization routines from
 {\sc lapack} \cite{lapack} to solve
equation (\ref{eq:xiseparada}) directly. However, within this
approach it is not simple to introduce explicitly the boundary
conditions in contrast to our GSF expansion approach for which it
is straightforward. Thus, our method allows us to obtain very
accurate results, even with a very few number of points in the
numerical grid. Nevertheless, since all the required integrals are
one--dimensional, we used a significant number of points ($\sim
10^4$), regardless of whether it was necessary.

Having solved the Sturmian equation and generated the basis set 
${\cal S}^r_i(\xi)$ for a chosen external parameter $E_s$, we can proceed to 
analyze convergence issues for the expansion (\ref{eq:expansionxi}) of the 
function $U(\xi)$. 
In Table \ref{table:convergenciaE} the basis size dependence of the energy 
value $E$, obtained by fixing the separation constant $A = 0.8117296$, 
is shown for two different sets. 
As explained in section \ref{subsubsec:radial}, the external parameter $E_s$ 
is an arbitrary energy; however, it is convenient to choose a value 
close to the true state energy. 
In a first calculation, we took the value $E_s = -1$ a.u. and  obtained the 
convergence sequence shown in the second column of 
Table \ref{table:convergenciaE} that leads to a state energy of 
$E = -1.1026$ a.u.. 
In a second, better, calculation we generate the radial GSF basis using as 
the external Sturmian energy, precisely this state energy, i.e., we set 
$E_s = -1.1026$ a.u.. 
In so doing, the sequence of energies obtained, listed in the third column of 
the table, converges very fast to the very accurate benchmark value.

\begin{table}[h]
\centering
\begin{tabular}{| c | l | l |}
\hline 
\rule{0pt}{1.1\normalbaselineskip}   
Basis Elements  &
~~$E$ (a.u.)  &  ~~$\tilde{E}$ (a.u.) \\
\hline
\textbf{$1$} & -1.0 & -1.1\\
\hline
\textbf{$3$} & -1.1 & -1.1026\\
\hline
\textbf{$6$} & -1.1024 & -1.1026340\\
\hline
\textbf{$9$} & -1.1026 & -1.1026346\\
\hline
Reference \cite{paperCN} & -1.1026342 & ~\\
\hline
\end{tabular}
\caption{Convergence of the energy $E$ in Eq.
(\ref{eq:xiseparada}) for fixed $A=0.8117296$ as a function of the
number of basis elements. The third column corresponds to the
energy $\tilde{E}$ obtained with an improved (recalculated)
basis.} \label{table:convergenciaE}
\end{table}

\begin{figure}[h]
\centering
\includegraphics[width=0.7\textwidth]{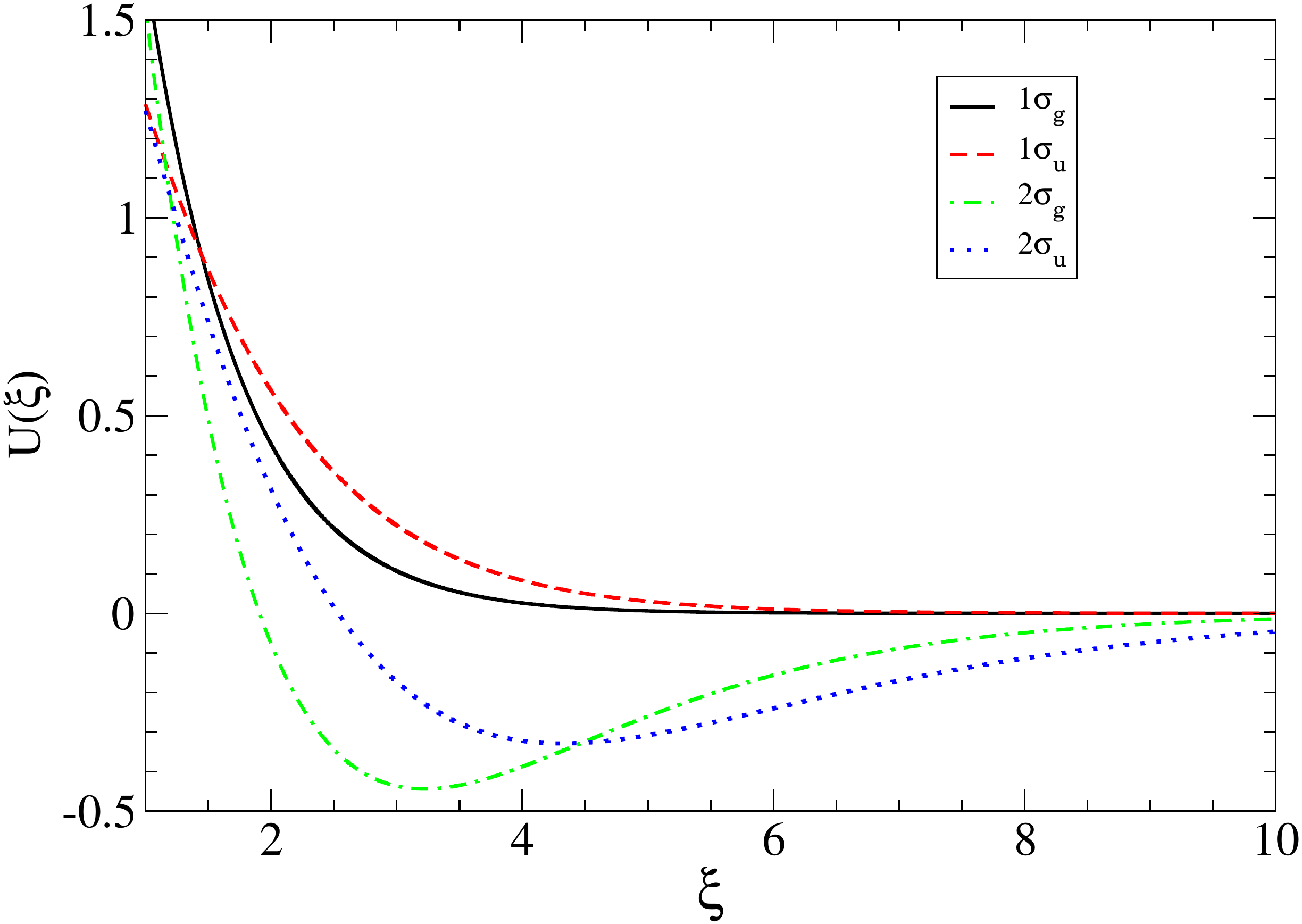}
\caption{The radial $U(\xi)$ solutions for the four lower energy
states of H$_{2}^{+}$.} \label{fig:4estadosxi}
\end{figure}

Once the eigenvalues equation is solved, the eigenvectors of
(\ref{eq:autovaloresradial}) provide the expansion coefficients
$d_i$, which build the radial function $U(\xi)$ through
(\ref{eq:expansionxi}). The converged result is shown in Figure
\ref{fig:4estadosxi}; the excited states will be discussed in the
next section.

The product of the angular and radial solutions $\Lambda(\eta)
U(\xi)$ gives, up to the azimuthal dependence, the wavefunction
which is best visualized by converting  the prolates $(\xi ,\eta
,\phi )$ into cartesian coordinates $(x,y,z)$ through
\begin{subequations}
\begin{eqnarray}
x &=& \frac{R}{2}\sqrt{(1-\eta^2)(\xi^2-1)} \cos(\phi)\\
y &=& \frac{R}{2}\sqrt{(1-\eta^2)(\xi^2-1)} \sin(\phi)\\
z &=& \frac{R}{2}\eta \xi\, .
\end{eqnarray}
\end{subequations}
In the top left panel of Figure \ref{fig:H+H+evsR} we show the
obtained $\psi_{1\sigma_g}$ for a fixed angle $\phi$ (for $m=0$
states, the results are symmetric respect to rotations over the
$z$ axis, and therefore, there is no dependence on the angle
$\phi$).

%%%%%%%%%%%%%%%%%%%%%%%%%%%%%%%%%%%%%%%%%%%%%%%%%%%%%%%%%%%%%%%%%%%%%%
\subsubsection{Internuclear distance dependence} %\hfill\\

\begin{figure}[h]
\centering
\includegraphics[width=0.7\textwidth]{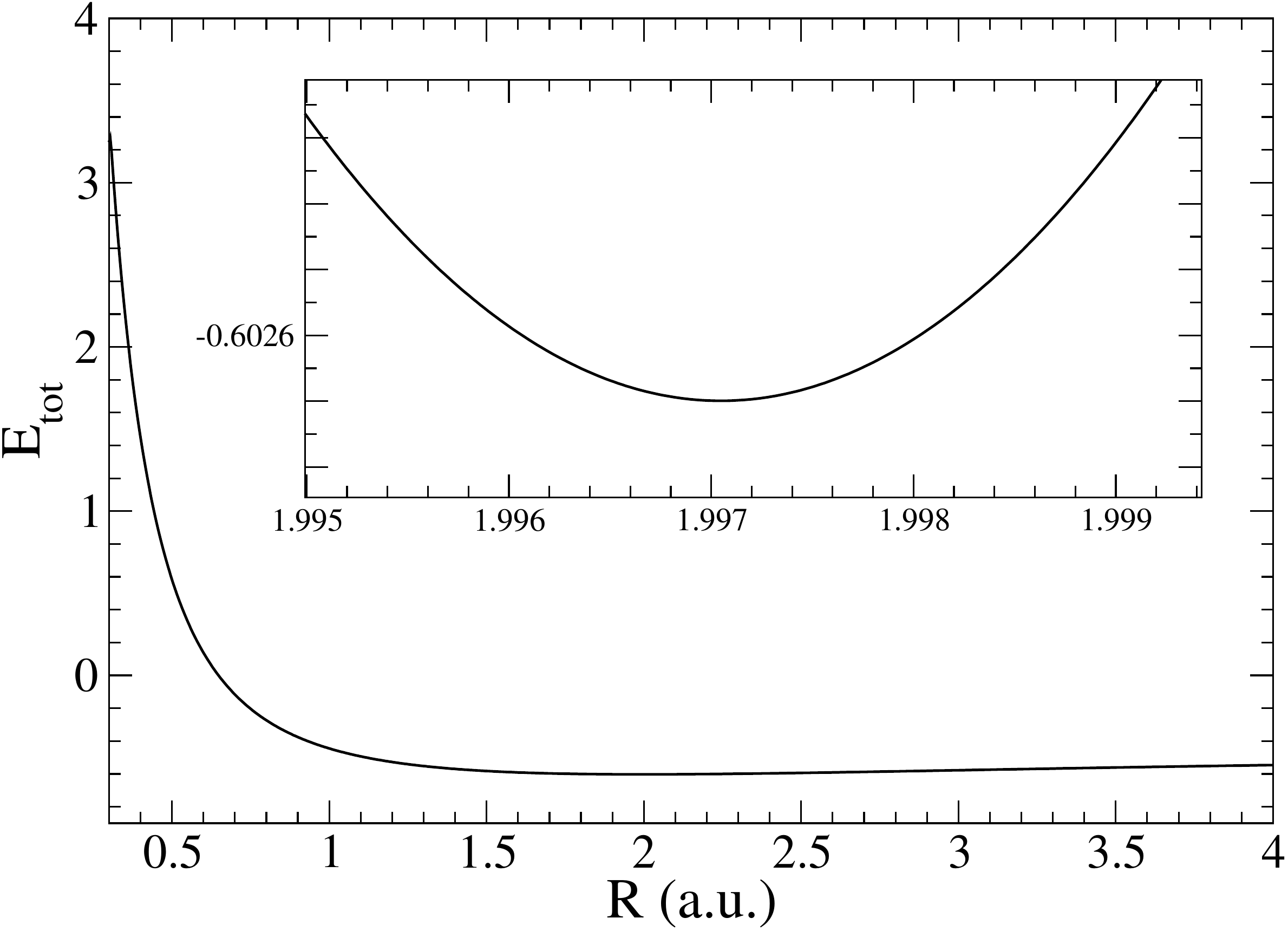}
\caption{Total energy  of the H$_{2}^{+}$ ground state as a
function of the internuclear distance $R$.} \label{fig:Rdepend}
\end{figure}

In the ground state results presented above we have fixed,
adopting the Born--Oppenheimer approximation, the internuclear
distance at $R=2$ a.u.
Calculations can be easily repeated by varying $R$, and in each
case, one obtains the total energy
\begin{equation}
E_{\mathrm{tot}}(R)= E(R) + \frac{1}{R} \, .
\end{equation}
The radial Sturmian functions should be generated through Eq.
(\ref{eq:sturxi}) in which one modifies $a_2=R(Z_1+Z_2)$ for each
$R$. This option can be taken but we found it convenient to use a
unique basis generated with a given value $a_{2s} = R_s (Z_1 +
Z_2) $;  except for very high internuclear distances $R$, we
simply took $R_s=2$ a.u.. In so doing, the use of the Sturmian
equation for the radial Schr\"odinger equation
(\ref{eq:xiseparada}) leads to a slightly modified Eq.
(\ref{eq:xiespansion}) and thus the supplementary matrix element
$\, (a_2 - a_{2s}) \, [\mathbf{{\cal N}^1}]_{ij} $ must be added
to matrix $\mathbf{N}$ defined by (\ref{Nmatrix}).
We have calculated the total energy for many values of $R$ taking 4 
angular and 6 radial basis functions, generated with a Sturmian 
energy $E_s=-1.1026340$, which is the energy value obtained for $R=2$ a.u. 
in the previous section.
Figure \ref{fig:Rdepend} presents the resulting energy $E_{tot}$ as a 
function of the internuclear distance. The inset allows one to see
a clear  minimum  at $R=1.99704$ a.u. At this equilibrium
distance (bond length) the corresponding  energy
$E_{\mathrm{tot}}=-0.602635$ a.u. is in agreement with the best
values given in the literature \cite{Requil}.

\begin{figure}[p]
%\vspace{-0.65cm}
\mbox{\begin{minipage}[t]{1.0\textwidth} \hspace*{-2cm}
\includegraphics[width=0.7\textwidth]{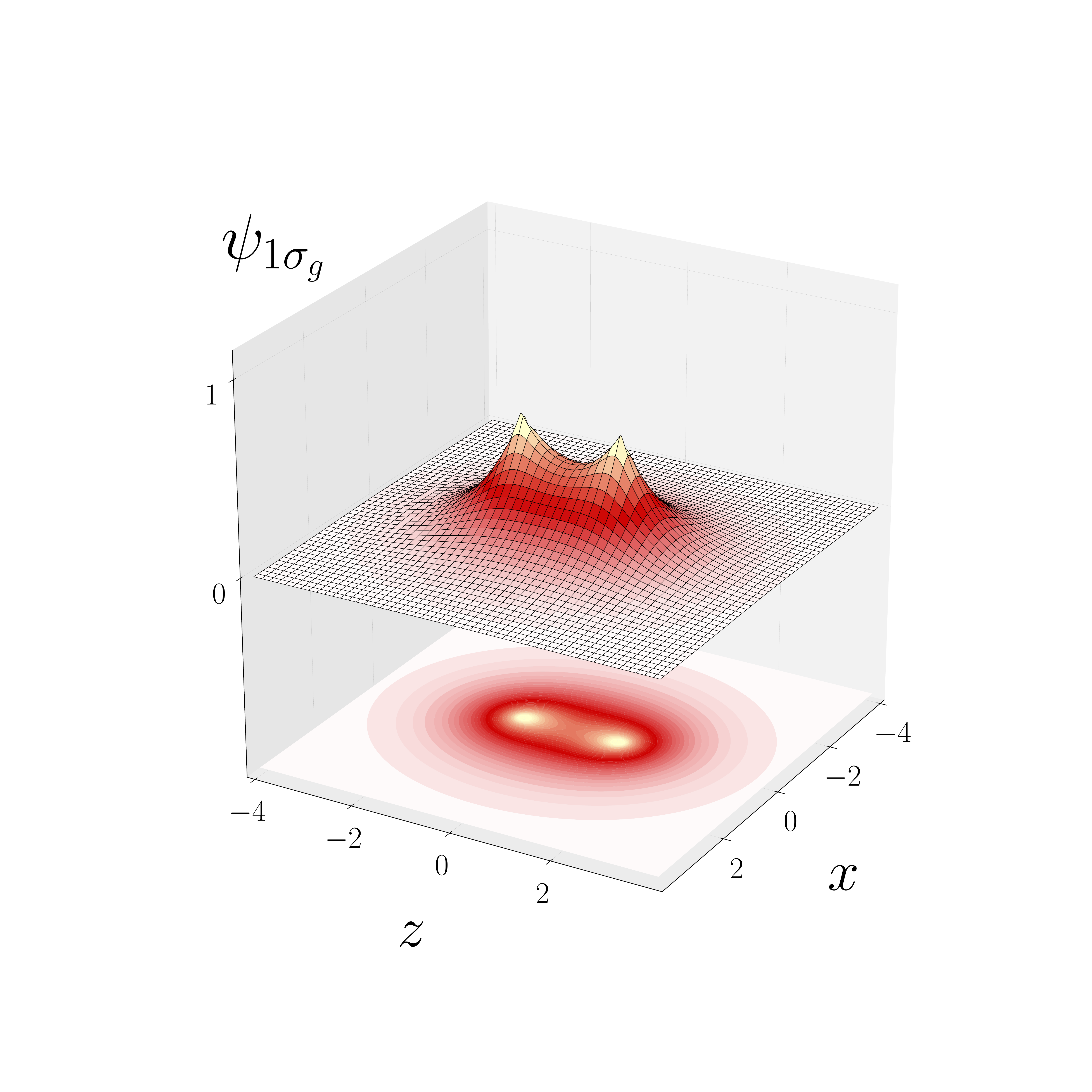}
\hspace*{-2cm}
\includegraphics[width=0.7\textwidth]{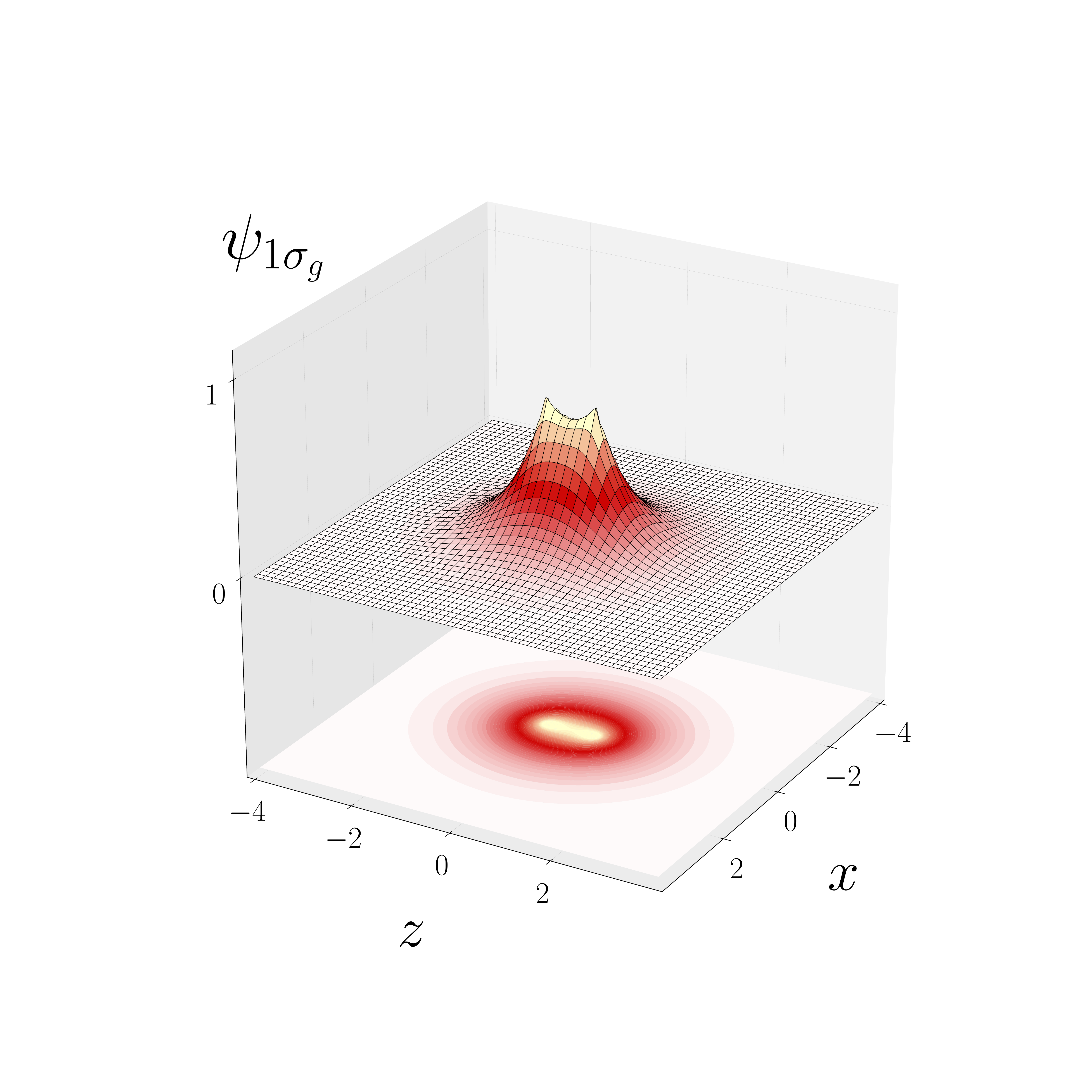}
\end{minipage}}
\mbox{\begin{minipage}[t]{1.0\textwidth} \vspace{-2.00cm}
\hspace*{-2cm}
\includegraphics[width=0.7\textwidth]{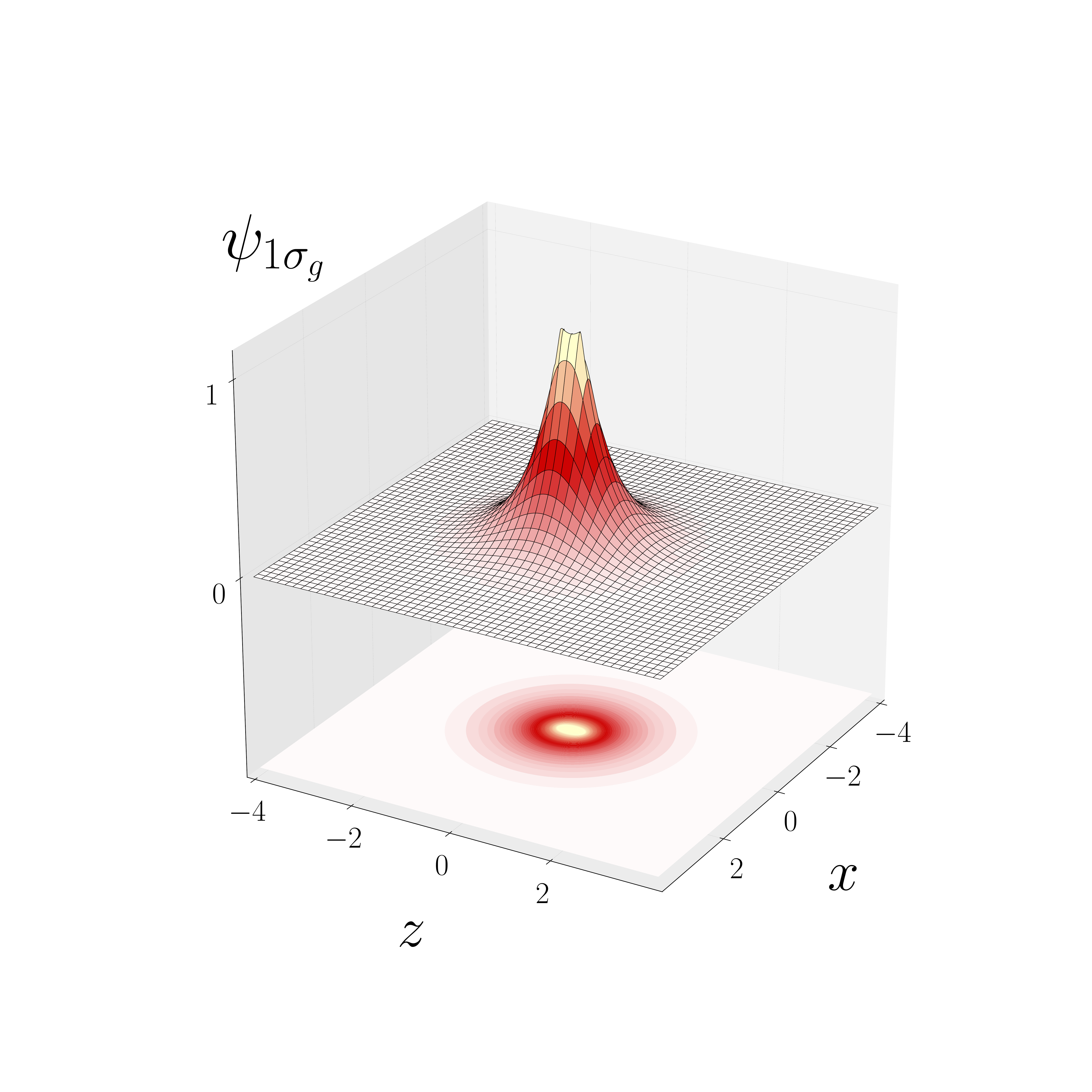}
\hspace*{-2cm}
\includegraphics[width=0.7\textwidth]{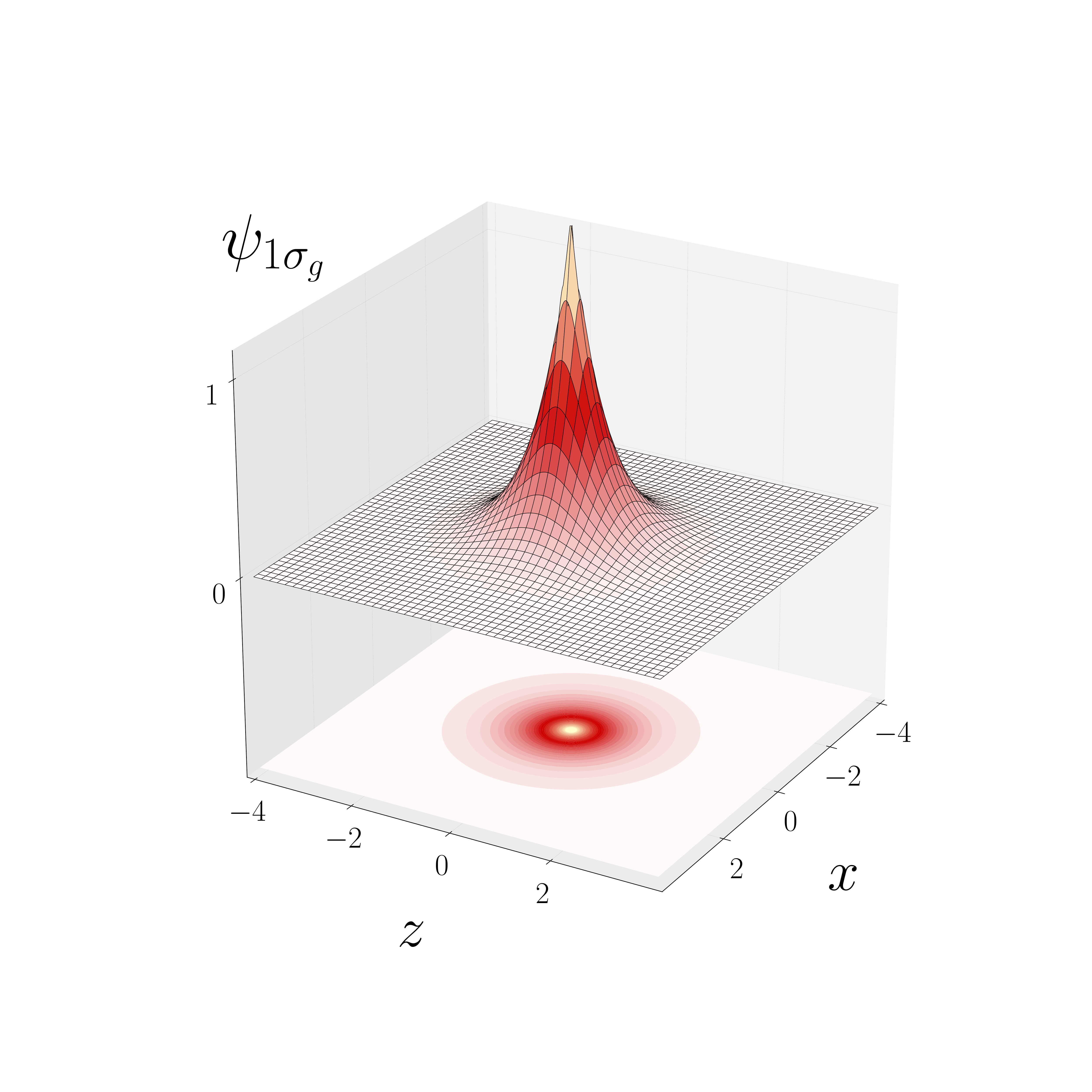}
\end{minipage}}
\caption{Wavefunction $\psi_{1\sigma_g}$ converted to cartesian
coordinates, for the H$_2^+$ ground state, calculated at four
different internuclear distances, moving
 from the molecular ion
H$_{2}^{+}$ to the atomic ion He$^+$: (top left) $R=2.0$ a.u.,
(top right) $R=1.0$ a.u., (bottom left) $R=0.4$ a.u., (bottom
right) $R=0.008$ a.u. To better appreciate the evolution  we have
renormalized the wavefunctions.} \label{fig:H+H+evsR}
\end{figure}

We challenged our computational method with energy calculations
considering very small internuclear distances $R$ for which, in
general, many numerical instabilities and errors arise. The energy
values displayed in Table \ref{table:Rs} demonstrate that our
Sturmian method remains robust for decreasing distances $R$, even
in the limit $R\rightarrow 0$, for which the solution corresponds
to the atomic ion  He$^{+}$ with energy $E_{He^+} = -Z^2/2 = -2$
a.u.. At the same time the ground state wavefunction should evolve
from a molecular to an atomic shape, that is to say from a density
of probability centered on the two nuclei to a hydrogenic single
center system. This transition from  molecular to atomic system as
the internuclear distance decreases is illustrated in
 Figure \ref{fig:H+H+evsR}.

\begin{table}[h]
\centering
\begin{tabular}{| l | l |}
\hline
\rule{0pt}{1.1\normalbaselineskip} 
 \centering R (a.u.) &    ~~$E$ (a.u.)  \\ \hline
2 & -1.1026340\\      \hline
1 & -1.4517823\\      \hline
0.4 & -1.800754\\     \hline
0.1 & -1.9782552\\    \hline
0.025 & -1.9984113\\  \hline
0.008 & -1.9998307\\  \hline
He$^{+}$ & -2.0\\
\hline
\end{tabular}
\caption{Ground state energy of the system $H+H+e^-$,
as a function of the internuclear distance $R$.}
\label{table:Rs}
\end{table}

\clearpage
%%%%%%%%%%%%%%%%%%%%%%%%%%%%%%%%%%%%%%%%%%%%%%%%%%%%%%%%%%%%%%%%%%%%%%%
\subsection{Iterative $1d$ method for some excited states of H$_2^+$}

By modifying the way the GSF basis functions are constructed, the
GSF spectral method allows one to obtain not only the ground state
but also excited and continuum states. To start with, let us look
at the first excited state $1\sigma_u$. For the generation of the
radial basis, it is necessary to choose an arbitrary Sturmian
energy as an external parameter. In a first, crude, approach we
take the same energy obtained for the ground state calculation
($E_s=-1.10263$ a.u. or, equivalently, $p_s=1.485015$). We
generate then three Sturmians for the angular basis (only odd
functions because of parity) and six radial Sturmian functions.
With these functions, we carry out the iteration procedure,
solving first the angular equation, obtaining the eigenvalue $A$.
This value is introduced as a parameter into the radial equation,
whose solutions produce a new scaled energy value $p$. As shown in
Table \ref{table:conv1su}, a very precise result with six
significant figures is obtained after only eight iteration steps.
However, as we discussed for the ground state, we can make the
whole calculation even better, choosing the Sturmian energy value
from the last convergence step ($p_s= 1.154791$, or
$E_s=-0.666771$ a.u.) and recalculating the radial basis. In so
doing,  the convergence is even faster, and only four iteration
steps are sufficient to reach the energy values given by Scott
\cite{Scott}.
\begin{table}[h]
\centering
\begin{tabular}{|c|c|c|c|c|}
\hline 
\rule{0pt}{1.1\normalbaselineskip} 
  Iteration  &  $p$ &  $E$ (a.u.) & $\tilde{p}$ &  $\tilde{E}$ (a.u.) \\
\hline 0 & 1.485015 & -1.10263  & 1.154791 & -0.666771  \\ \hline
2 & 1.175548 & -0.690957 & 1.155444 & -0.667525   \\ \hline 4 &
1.155869 & -0.668017 & 1.155451 & -0.667534  \\ \hline 6 &
1.154793 & -0.666773 &~&\\ \hline 8 & 1.154791 & -0.666771 &~&\\
\hline
Reference \cite{Scott} & 1.155452 & -0.667534 &~& \\
\hline
\end{tabular}
\caption{Convergence of $p$ and energy $E$ for the first H$_2^+$
excited state $1\sigma_u$. The fifth column corresponds to the
energy $\tilde{E}$ obtained with an improved (recalculated)
basis.} \label{table:conv1su}
\end{table}

The same procedure is repeated for the generation of other excited
states, such as  $2\sigma_g$ and $2\sigma_u$. In Table
\ref{table:excited1d}  the energy results obtained with our
iterative method are displayed and compare very favorably with the
results obtained by Bian \cite{paperCN}. Note that the latter
coincide, up to the eighth decimal with those of Madsen and Peek
\cite{H2+eigenpar}.
\begin{table}[h]
\begin{center}
\begin{tabular}{| c | c | c | c | c |}
\hline 
\rule{0pt}{1.1\normalbaselineskip} 
State  &  $A$ & $E$ (a.u.) &  $\tilde{E}$ (a.u.) &
$E$ (a.u.) Ref.  \cite{paperCN}\\
\hline
\textbf{$1\sigma_{g}$} & 0.8117 & -1.102 & -1.1026340 & -1.10263421\\
\hline
\textbf{$1\sigma_{u}$} & -1.8689 & -0.667 & -0.6675338 & -0.66753439\\
\hline
\textbf{$2\sigma_{g}$} & 0.2484 & -0.3 & -0.36081 & -0.36086488\\
\hline
\textbf{$2\sigma_{u}$} &  -1.69179  &-0.25 & -0.25535 & -0.25541317\\
\hline
\end{tabular}
\caption{Parameter $A$ and energies $E$ of the lowest energy
states of H$_{2}^{+}$ calculated with our iterative GSF method.
The fourth column corresponds to the energy $\tilde{E}$ obtained
with an improved (recalculated) basis. The last column reports the
energy values found by Bian \cite{paperCN}.}
\label{table:excited1d}
\end{center}
\end{table}

\begin{figure}[h!]
\mbox{\begin{minipage}[t]{1.0\textwidth}
\hspace*{-2cm}
\includegraphics[width=0.7\textwidth]{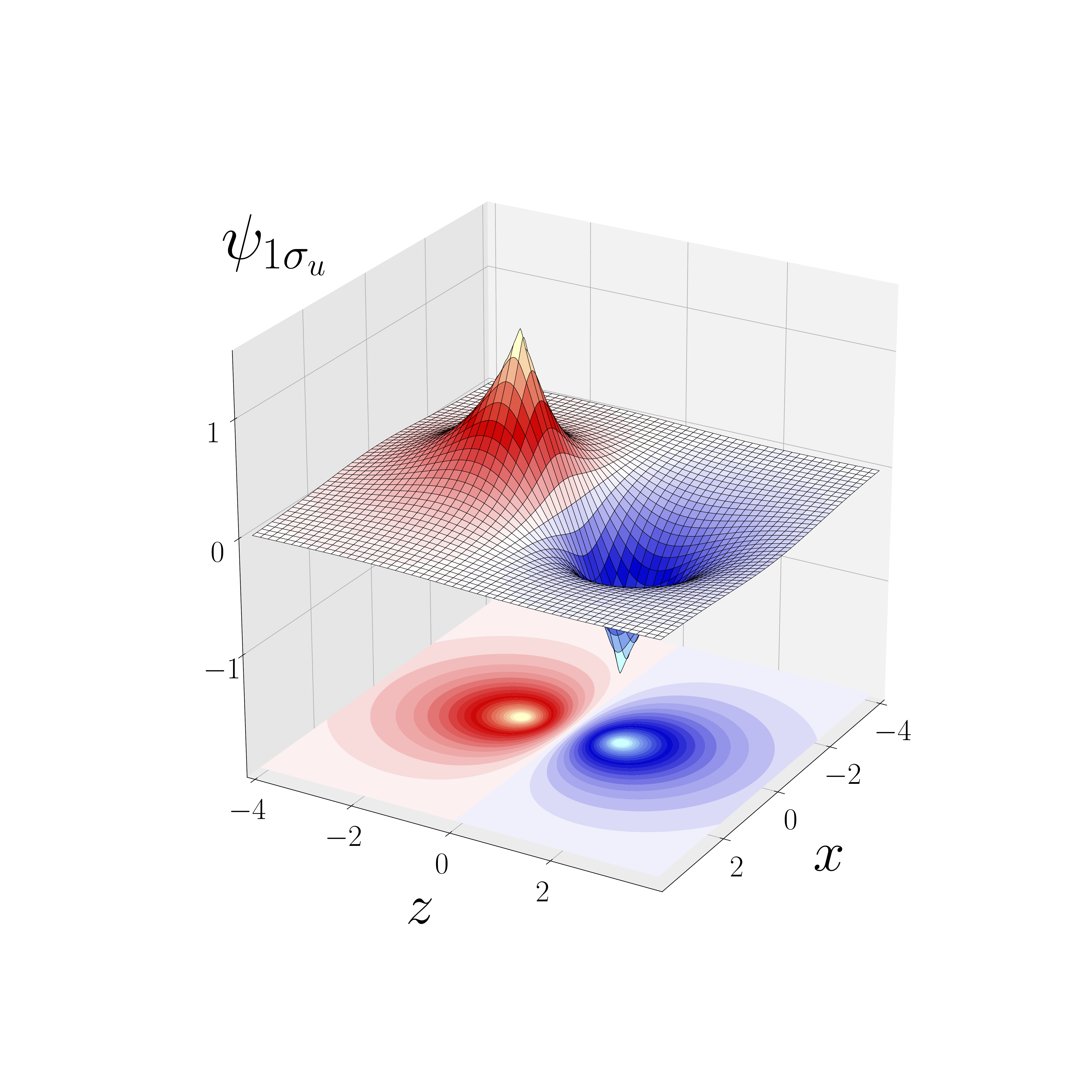}
\hspace*{-2cm}
\includegraphics[width=0.7\textwidth]{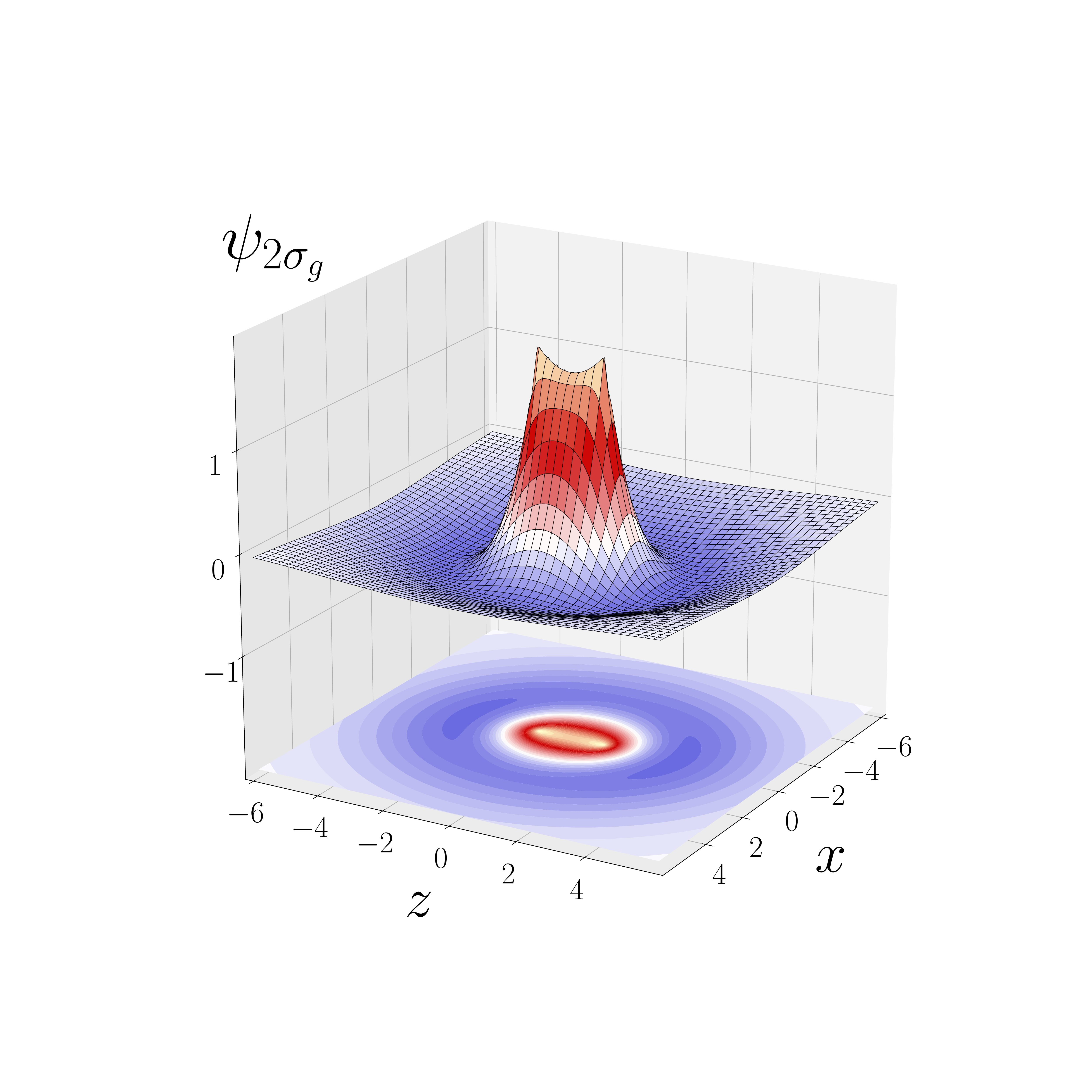}
\end{minipage}}
\mbox{\begin{minipage}[t]{1.0\textwidth}
\vspace{-1.95cm}
\hspace*{0.225\textwidth}
\includegraphics[width=0.7\textwidth]{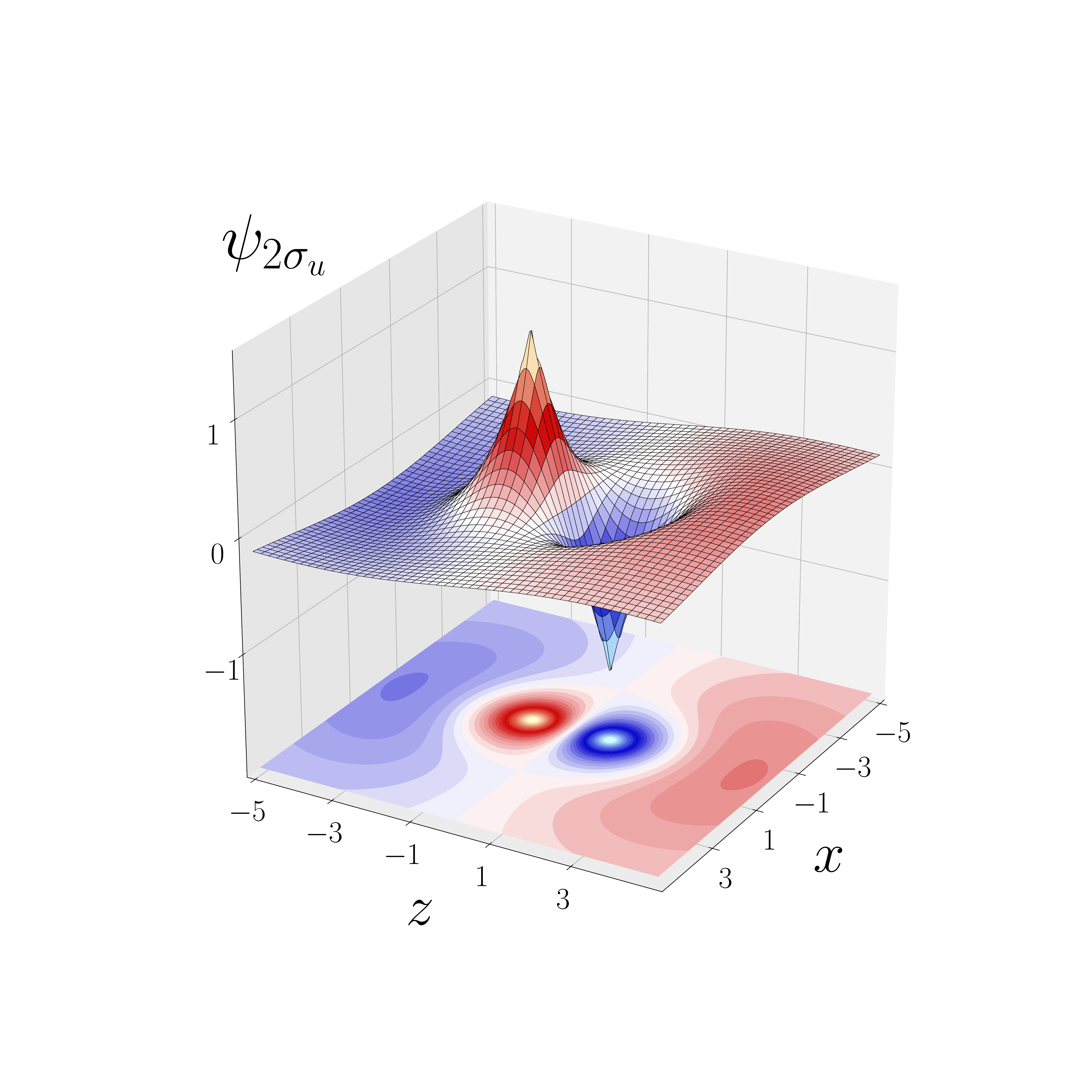}
\end{minipage}}
\caption{Wavefunctions for the first excited states $1\sigma_u$,
$2\sigma_g$ and $2\sigma_u$ of H$_2^+$.} \label{fig:1sxyu}
\end{figure}

The radial $U(\xi)$ and the angular $\Lambda(\eta)$ solutions of
the four lowest states of H$_{2}^{+}$ are shown, respectively, in
Figures \ref{fig:4estadoseta} and  \ref{fig:4estadosxi}. The total
wavefunctions for the excited states $1\sigma_u$, $2\sigma_g$ and
$2\sigma_u$ are shown in Figure \ref{fig:1sxyu} as a function of
the cartesian coordinates $(x,z)$. We recall that the density is
invariant under rotations around the $z$ axis.

\clearpage
%%%%%%%%%%%%%%%%%%%%%%%%%%%%%%%%%%%%%%%%%%%%%%%%%%%%%%%%%%%%%%%%%%%%%
\subsection{Iterative $1d$ method for the asymmetric molecular ions
HHe$^{+2}$ and HLi$^{+3}$}

We apply now our GSF approach to other monoelectronic diatomic
systems, such as the HHe$^{+2}$ and HLi$^{+3}$ molecular ions.
For these heteronuclear ions, $Z_1\neq Z_2$ and thus $a_1
\neq 0$. In order to compare with other sample results published
in the literature, we have kept the internuclear distance fixed at
$R=4$ a.u. (for HHe$^{+2}$ the equilibrium value is around
$R=3.89$ a.u.).
These molecular ions are no longer symmetric along the $z=0$ axis, 
so that the angular representation in Legendre polynomials requires 
many more elements than the -- symmetric --  H$_2^+$ case. 
This said, the computational cost is not significantly increased since all the 
angular integrals  are analytical. For the HHe$^{+2}$ molecular ion, 
we used 20 angular and 6 radial basis functions. We performed an initial 
calculation choosing the Sturmian energy $E_s=-3.0$ a.u., obtaining a 
ground state energy $E=-2.25060$, a value that was then recycled as the new $E_s$.
For the HLi$^{+3}$ molecular ion, we used 100 angular and 6 radial basis 
functions, starting with an initial guess $E_s=-5.0$ a.u., 
obtaining $E=-4.74968$ a.u. then recycled as the new $E_s$.

\begin{figure}[h!]
%\vspace{-0.65cm}
\mbox{\begin{minipage}[t]{1.0\textwidth} \hspace*{-2cm}
\includegraphics[width=0.7\textwidth]{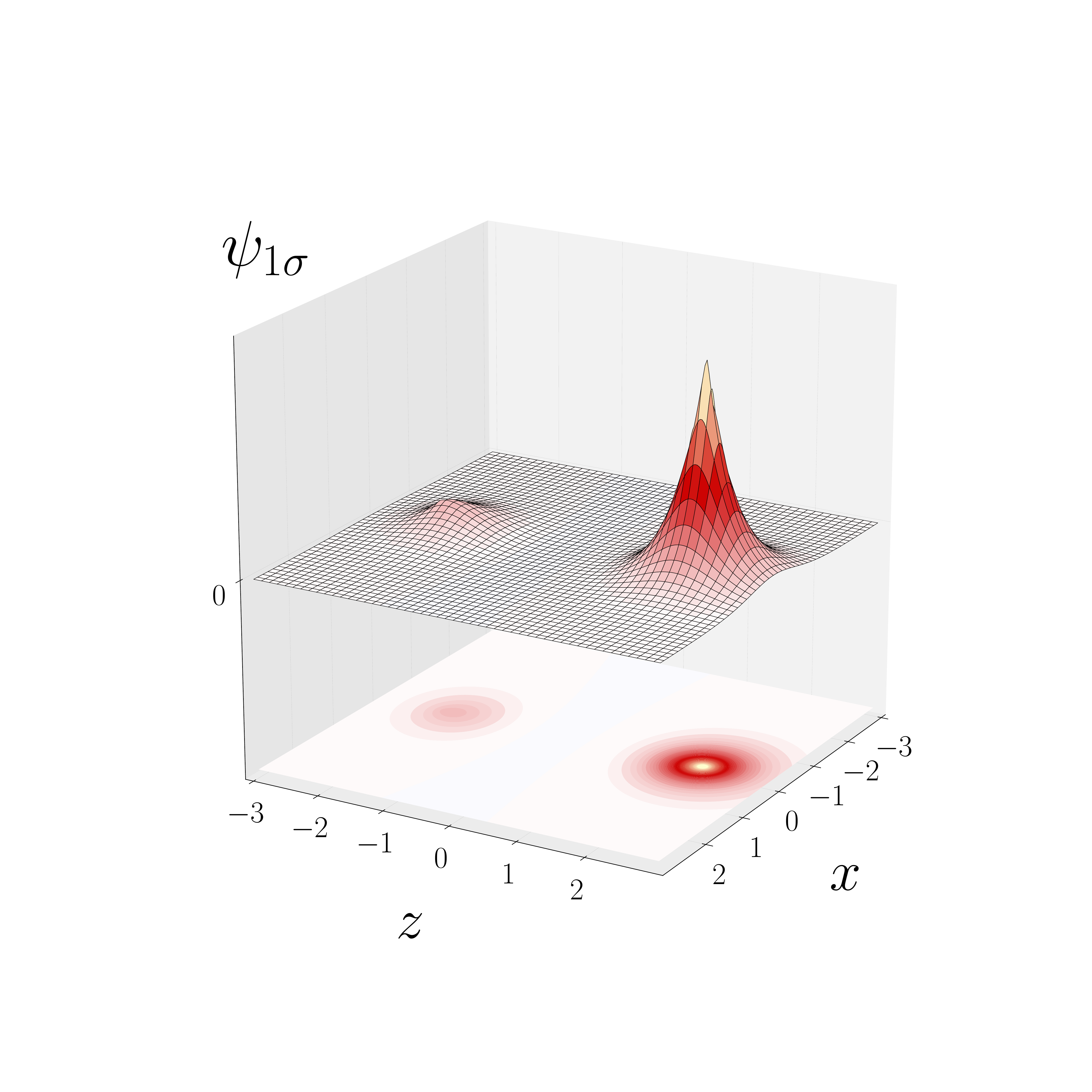}
\hspace*{-2cm}
\includegraphics[width=0.7\textwidth]{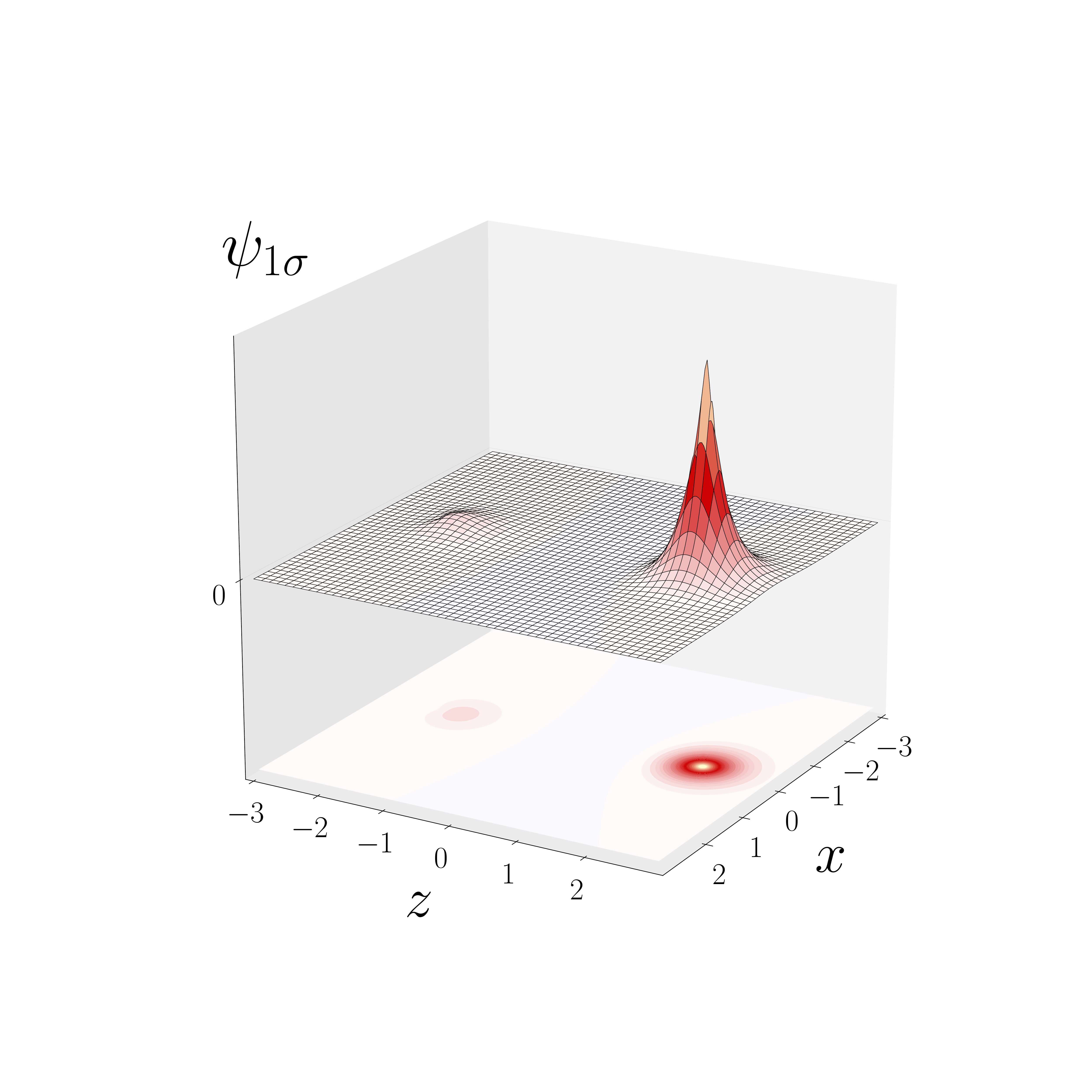}
\end{minipage}}
\caption{Unnormalized ground state wavefunctions 
$1\sigma$ of HHe$^{+2}$ (left)
and HLi$^{+3}$ (right), assuming an internuclear distance $R=4$
a.u..} \label{fig:1sgHLi}
\end{figure}

The ground state wavefunctions of the heteronuclear molecular ions
are shown in Figure \ref{fig:1sgHLi}. The distribution of the
electron density is now clearly asymmetric, the logical shift
towards the nucleus with larger charge being more evident as the
Coulomb attraction increases. The shape of the wavefunction
acquires more and more an atomic--like form centered on the
heavier nucleus with only relatively small values close to the
hydrogen nucleus. These features will obviously strongly depend on
the internuclear distance, here fixed at $R=4$ a.u..

Table \ref{table:HHe} displays the calculated ground state
energies, whose absolute value increases approximately as
$Z_2^2/2$ with $Z_2$ the charge of the heavier nucleus. The
efficiency of our method can be appreciated by giving a few
numbers of other methods. The results given by Avery {\it et al.}
\cite{Avery:09} were calculated with 10 basis elements (Coulomb
Sturmian functions) for each nucleus. Kereselidze {\it et al.}
\cite{Rusos} used 10 basis functions per nucleus (Coulomb Sturmian
in prolate spheroidal coordinates). Xue--Bin Bian \cite{paperCN}
employed an imaginary--time--propagation method based on a
Crank--Nicolson scheme to solve the separate equations, using 20
B--splines of order 7 to solve the radial equation, and 80
B--splines of order 7 for the angular part. Campos  {\it et al.}
\cite{Campos} used 22 functions per coordinate. The aim of our
calculation here was not to obtain very high accuracies, but
rather to demonstrate that our simple and versatile method is
computationally more efficient when compared to other
approaches. If desired, we can achieve even better energy
accuracies by improving the employed numerical methods (number of
points or the finite differences order) or by tuning the 
generating potential as to optimize the GSF basis set.
\begin{table}[h]
\centering
\begin{tabular}{| c  | c | c | c |}
\hline
\rule{0pt}{1.1\normalbaselineskip} 
   &  $E$ $1\sigma_g$  H$_{2}^{+}$ (a.u.)  &  $E$ $1\sigma$
   HHe$^{+2}$  (a.u.) &  $E$ $1\sigma$ HLi$^{+3}$ (a.u.) \\
\hline
This work & -1.1026340 & -2.2506056 & -4.7501126\\
\hline
Avery \cite{Avery:09} & -1.10220 & - & -4.75011\\
\hline
Kereselidze \cite{Rusos} & -1.102614 & - & -4.750111\\
\hline
Bian \cite{paperCN} & -1.1026342 & -2.2506054 & -\\
\hline
Campos \cite{Campos} & - & -2.2506054 & -4.7501118\\
\hline
\end{tabular}
\caption{Ground state energies for the monoelectronic molecular
ions: H$_{2}^{+}$, assuming an internuclear distance
$R=2$ a.u., and  HHe$^{+2}$ and HLi$^{+3}$, assuming an
internuclear distance $R=4$ a.u.}.
\label{table:HHe}
\end{table}

%%%%%%%%%%%%%%%%%%%%%%%%%%%%%%%%%%%%%%%%%%%%%%%%%%%%%%%%%%%%%%%%%%%%%%
\section{Direct $2d$ method
for the ground and excited states of H$_{2}^{+}$}

Although we found excellent results with the iterative method, we
wish to exploit the full advantages of the spectral method which
allows one to obtain many states in one shot. The direct
diagonalization of a $2d$ Hamiltonian is generally very costly
from the computational point of view. Within the finite
differences framework, and taking into account that every
coordinate has to be represented by hundreds of points, the matrix
becomes huge and is intractable. A spectral method can reduce
significantly the size of the Hamiltonian matrix to diagonalize,
but computationally it still represents a hard task. Within the
GSF method, the size of the matrices are reduced even more, since
the appropriate physical behavior is explicitly introduced in the
basis set. In this manner, the numerical treatment is optimized.

The use of expansion (\ref{expansion2d}) on a two--dimensional
basis ${\rm S}_{ij}(\xi,\eta)$ transforms the Schr\"odinger
equation into an equation (\ref{eq:H2dstur}) where  all the
derivatives have been removed and replaced  by simple expressions.
Moreover, since the basis functions are optimized, the size of the
basis is very small. For example, in our calculations, we
introduced only 18 functions (3 angular $S^a_j(\eta)$ and 6 radial
${\cal S}^r_i(\xi)$). Finally, the direct diagonalization of this
small matrix produces, as a result, 18 states simultaneously
without the need to perform separate iterations for each state.

We have applied our GSF direct $2d$ method to the benchmark ion
H$_{2}^{+}$, again taking $R=2$ a.u.. With only one
diagonalization we obtained the energy values displayed in Table
\ref{table:excited2D}. They compare very well with the results of
Madsen and Peek \cite{H2+eigenpar}, following their states
notation. We should point out that our aim here was to produce all
the levels at the same time without a focus on a single state. To
generate the Sturmian basis we chose here the energy value
$E_{s}=$-0.2 a.u. which is clearly quite different from the ground
state energy; it is an acceptable compromise that leads to a good
precision for the whole set of presented molecular states. The
table shows that  it is possible to obtain excellent results, in
particular for the lower states, at a rather small computational
cost. If one wishes to improve the energy accuracy for one
particular state, a different Sturmian energy $E_s$ closer to this
state energy should be chosen, as was shown in the $1d$ method.
Since the generation of a new Sturmian basis requires
one--dimensional calculations and the $2d$ matrix only has a few
dozen of elements, this further optimization procedure is rather
inexpensive.

\begin{table}[h]
\begin{center}
\begin{tabular}{| c | l | l |}
\hline
\rule{0pt}{1.1\normalbaselineskip} 
  State  &  ~~$E$ (a.u.) &
~~$E$ (a.u.) Ref. \cite{H2+eigenpar}\\
\hline
\textbf{$1S\sigma_{g}$} & -1.102630 & -1.10263421\\
\hline
\textbf{$2P\sigma_{u}$} & -0.66753431 &  -0.66753439\\
\hline
\textbf{$2S\sigma_{g}$} & -0.360863 & -0.36086488\\
\hline
\textbf{$3P\sigma_{u}$} & -0.25541312 & -0.25541317\\
\hline
\textbf{$3D\sigma_{g}$} & -0.2357775 & -0.23577763\\
\hline
\textbf{$3S\sigma_{g}$} & -0.1776 &  -0.17768105\\
\hline
\textbf{$4P\sigma_{u}$} & -0.133 & -0.13731293\\
\hline
\end{tabular}
\caption{Energies of seven energy states of H$_{2}^{+}$, obtained
with the GSF direct $2d$ method. The third column indicates the
results of Madsen and Peek \cite{H2+eigenpar}. Both were obtained
for a fixed internuclear distance $R=2$ a.u..}
\label{table:excited2D}
\end{center}
\end{table}

%%%%%%%%%%%%%%%%%%%%%%%%%%%%%%%%%%%%%%%%%%%%%%%%%%%%%%%%%%%%%%%%%%%%%%
\section{CONCLUSION}
\label{sec:conclusion}

The spectral method, based on Generalized Sturmian Functions, has
been here extended, to allow its use in prolate spheroidal
coordinates which should provide, in principle, the most effective
framework to treat diatomic molecular systems. We developed and
implemented two different numerical methods for the calculation of
the molecular structure of monoelectronic molecular ions.

The first one consists in separating the Schr\"odinger equation in
one angular and one radial equations, coupled through the energy
and a coupling parameter. The equations are solved alternately,
fixing the energy and the coupling parameter in each case, and
after a few iterations, these parameters converged to the final
values. The advantage of using GSF is twofold. On the one hand, it
allows one the replacement of most of the Hamiltonian calculations
by a simple expression thus substantially reducing the complexity
of the calculation at any iteration step. On the other hand, the
GSF method is based in the valuable property that the right
boundary conditions are enforced onto the basis functions.
Therefore, the size of the basis is minimal, turning the method in
a very efficient procedure that produces ground and excited states
of high quality.

The second method also uses GSF, and the angular and radial basis
sets are generated in the same way as in the first one. Then, a
two--dimensional basis set is constructed, and the Schr\"odinger
equation solution becomes a $2d$ generalized eigenvalues problem.
Since the basis elements have the correct boundary conditions, the
size of the basis is very small, and the diagonalization is not a
costly procedure. This direct $2d$ method does not require any
iteration and a single calculation yields -- simultaneously --
many molecular states. Very good results can be obtained already
with small basis set size.
Both methods are computational efficient, but a quantitative comparison 
is not appropriate. 
Indeed, in the $1d$ iterative method the GSF basis is generated as to focus 
on one particular state and great accuracy can be achieved. 
In contrast, the direct $2d$ method uses the same GSF basis to obtain a set of 
orthogonal bound states, and thus provides richer results albeit of relatively 
inferior accuracy. 
Besides, the spectrum obtained by diagonalization may include discretized 
states of the continuum which can be useful for collision studies. 
In other words, one may state that the $1d$ iteration
method is optimal to focus on a spectific state while the $2d$ method provides a
global view of the spectrum.

As a first step towards  the extension of the GSF method to
diatomic molecules, we have presented here an investigation of
molecular ions having only one electron. We calculated the ground
and excited states of the molecular hydrogen ion H$_{2}^{+}$, in
excellent agreement  with benchmark results (7 significant figures
in the case of the ground state). We also studied heteronuclear
molecular ions, like HHe$^{+2}$ and HLi$^{+3}$, with again
excellent results. The method proved to be robust over a wide
range of internuclear distances $R$, including in the notoriously
difficult atomic limit.

The whole numerical investigation gives us confidence in
our implementation of the GSF method in prolate spheroidal
coordinates, as to contemplate exploring the continuous part of of
the spectrum. As demonstrated for atomic systems, the advantages
of the GSF spectral method are more evident in the treatment of
collision problems. In this case, the continuum Sturmian basis
elements are generated with a positive energy parameter $E_s$ and
one imposes appropriate scattering boundary conditions. As a
consequence, the basis needs to solve the Schr\"odinger equation
only in the interaction region. Scattering problems involving one
or two electrons in the continuum can then be treated efficiently
with compact bases
\cite{Gasaneo:13,HeDPI,Ambrosio2016,Ambrosio2017}. The same
arguments apply to diatomic molecular systems, and we plan to
extend the present investigation in prolate spheroidal coordinates
to scattering problems such as single or double ionization by
photon or electron impact. First we will examine the
single continuum by studying the single photoionization of the
benchmark one--electron molecular ion H$_2^+$; then, we will move
to the more challenging two--electron correlated case, by
investigating single and double ionization processes on H$_2$ and
on quasi two--electron targets like N$_2$ as done for example in
Ref. \cite{Chuka2012,Bulychev2013}.

%%%%%%%%%%%%%%%%%%%%%%%%%%%%%%%%%%%%%%%%%%%%%%%%%%%%%%%%%%%%%%%%%%%%%%
\section{FOUNDING INFORMATION}

DM gratefully acknowledge the financial support from the following Argentine
institutions: Consejo Nacional de Investigaciones Cient\'{\i}ficas y
T\'ecnicas (CONICET), PIP 11220130100607, Agencia Nacional de Promoci\'on
Cient\'{\i}fica y Tecnol\'ogica (ANPCyT) PICT--2017--2945,
and Universidad de Buenos Aires UBACyT 20020170100727BA.

%%%%%%%%%%%%%%%%%%%%%%%%%%%%%%%%%%%%%%%%%%%%%%%%%%%%%%%%%%%%%%

%\pagebreak
%%%%%%%%%%%%%%%%%%%%%%%%%%%%%%%%%%%%%%%%%%%%%%%%%%%%%%%%%%%%%%
\end{document}